%% file: Accelerating_VFL.tex
\documentclass[lettersize, journal]{IEEEtran}
\makeatletter
\newif\if@restonecol
\makeatother

\usepackage{cite}
\usepackage[utf8]{inputenc}
\usepackage{graphicx}
\usepackage{amssymb}

\usepackage{hyperref}
\usepackage{multirow}
\usepackage[utf8]{inputenc}
\usepackage[ruled, lined, longend]{algorithm2e}
\usepackage{amsmath,multicol}
\usepackage{subcaption,textcomp,balance}

\usepackage{booktabs}
\usepackage{multirow}
\usepackage{color,xcolor}

\usepackage{changes}

\usepackage{array}
\newcolumntype{L}[1]{>{\raggedright\let\newline\\\arraybackslash\hspace{0pt}}m{#1}}
\newcolumntype{C}[1]{>{\centering\let\newline\\\arraybackslash\hspace{0pt}}m{#1}}
\newcolumntype{R}[1]{>{\raggedleft\let\newline\\\arraybackslash\hspace{0pt}}m{#1}}

\definecolor{red}{rgb}{0.8,0,0}

\begin{document}


\title{Accelerating Vertical Federated Learning}

\author{Dongqi~Cai,
        Tao~Fan,~\IEEEmembership{Member,~IEEE},
        Yan~Kang,
        Lixin~Fan,~\IEEEmembership{Member,~IEEE},
        Mengwei~Xu,~\IEEEmembership{Member,~IEEE},\\
        Shangguang~Wang,~\IEEEmembership{Senior Member,~IEEE},
        and~Qiang~Yang,~\IEEEmembership{Fellow,~IEEE}
\thanks{L. Fan and M. Xu are the corresponding authors. Email:  lixinfan@webank.com, mwx@bupt.edu.cn.}
\thanks{D. Cai, M. Xu and S.Wang are with State Key Laboratory of Networking and Switching Technology, Beijing University of Posts and Telecommunications, Beijing, China. L. Fan, T. Fan, Y. Kang are with Department of Artificial Intelligence, Webank, Shenzhen, China. Q. Yang is with both Webank and Hong Kong University of Science and Technology}}



\maketitle

\maketitle

\input{abstract}

%
\IEEEpeerreviewmaketitle

\input{introduction}

\input{motivation}

\input{method}

\input{experiments}

\input{related}

\input{conclusion}




\ifCLASSOPTIONcaptionsoff
  \newpage
\fi



%




\bibliographystyle{plain}
\bibliography{cdq}

\input{photos}

\end{document}

%% file: abstract.tex
\begin{abstract}
    Privacy, security and data governance constraints rule out a brute force process in the integration of cross-silo data, which inherits the development of the Internet of Things.
    Federated learning is proposed to ensure that all parties can collaboratively complete the training task while the data is not out of the local. 
    Vertical federated learning is a specialization of federated learning for distributed features. 
    To preserve privacy, homomorphic encryption is applied to enable encrypted operations without decryption. 
    Nevertheless, together with a robust security guarantee, homomorphic encryption brings extra communication and computation overhead. In this paper, we analyze the current bottlenecks of vertical federated learning under homomorphic encryption comprehensively and numerically. 
    We propose a straggler-resilient and computation-efficient accelerating system that reduces the communication overhead in heterogeneous scenarios by 65.26\% at most and reduces the computation overhead caused by homomorphic encryption by 40.66\% at most. Our system can improve the robustness and efficiency of the current vertical federated learning framework without loss of security.
\end{abstract}


%% file: introduction.tex
\section{Introduction}
%
%
%
%

\IEEEPARstart{D}{ata} fuels the growing popularity of the Internet of Things. 
With the ubiquitous deployment of dedicated and multi-purpose sensors, collecting real-time information about our environment is becoming more convenient.
However, due to the specific limitations of dedicated sensors, information collected by different industries tends to have divergent feature dimensions.
This divergence makes cross-silo training and inferring challenging. 
The traditional method is to establish a centralized data center led by relevant authorities. 
Although this can break the data isolation, it is not feasible today. 
Because of the tremendous value of this real-time information, such convenient access to monitor surroundings makes people concerned about the leakage and abuse of their private and sensitive data. 
Apart from the individual privacy concern, the implementation of user privacy laws such as GDPR~\cite{2018Regulation} has set a strict limit on the usage of the collected data. 
Therefore, cross-silo data sharing becomes challenging, which hinders the development of the Internet of Things.

To solve this dilemma, federated learning (FL)~\cite{1} was proposed by Google in 2016.
FL ensures that all parties can collaboratively complete the training task while the data is not out of the local, which is an ideal framework to alleviate privacy concerns. 
However, most FL research focuses on the sample partitioned scenario~\cite{Yang2019, Wang2022}, where all cooperators share the same feature dimension. 
The feature partitioned federated learning is rarely explored in literature, which is an equally important issue in real industrial scenarios, such as recommender system~\cite{Hu2019}, credit evaluation~\cite{Liu2019}, etc. 
\input{fig-fl}
Vertical federated learning (VFL) is a specialization of federated learning for distributed features~\cite{Hardy2017}.
As shown in Fig.~\ref{fig-fl}, economic and medical institutions have divergent feature dimension. 
They build a federated model together without exchanging raw data. 
To further preserve privacy, federated model is in safekeeping by a third-party authority, which prohibits each institution from getting model structure of others.
Unlike conventional FL (or noted horizontal FL) for distributed samples, which has been researched a lot due to its continuous line with traditional distributed learning, VFL is under-studied for the lack of datasets and benchmarks. 
Apart from the lack of related infrastructure, communication redundancy in VFL is considerably larger than in conventional FL. 
Because not only gradients are transmitted during cooperation, intermediate results such as labels and loss are needed to be exchanged due to distributed features~\cite{Yang2019,Wang2022}.  

Homomorphic encryption (HE)~\cite{Zhang2020}, a common privacy-preserving technology, is applied to provide secure protection of exchanged information. 
Apart from HE, differential privacy~\cite{dp}, quantization~\cite{reisizadeh2020fedpaq}, compression~\cite{sattler2019robust}, and coded computing~\cite{ng2021hierarchical} can be applied to preserve the privacy as well.
However, they are vulnerable to aggressive Bayesian restoration attacks due to their randomization privacy protection~\cite{triastcyn2020bayesian,bu2020deep,yang2020federated,zhu2019deep}.
Considering that cooperators in VFL training need to have fruitful data features for effective collaborative training, they tend to be competitive enterprises in certain fields, such as industry, finance, medicine, etc.
Such a potential adversarial relationship resorts to the most strict privacy preservation for these precious data.
HE ensures security by allowing encrypted operations without decryption and thus providing cryptographic privacy protection. 
Consequently, HE is studied as our focus of privacy preservation.




Nonetheless, in addition to the robust security guarantee, HE leads to two orders of magnitude size inflation of the original data after encryption, which burdens the communication and incurs massive delays. 
For example, a few-minute local training task will take dozens of minutes or even hours to complete in VFL under homomorphic encryption.
This extra delay, together with the common heterogeneity~\cite{Luo2020} existing in cooperation learning, damages the runtime performance of VFL. 
To make matters worse, unlike conventional FL, parties in vertical federated learning can not compute gradients locally due to the lack of labels. 
To compute the gradient, they need to exchange the encrypted intermediate results. 
Computational costs under homomorphic encryption are much higher than that under plain text.
Huge computation overhead further slows down vertical federated learning under homomorphic encryption.



In this paper, we analyze the main bottlenecks of current vertical federated learning numerically and implement a practical and efficient accelerating system under FATE
\footnote{Federated AI Technology Enabler (FATE) is an open-source industrial federated learning framework, which has been widely applied in the financial and medical fields. Git repository: https://github.com/FederatedAI/FATE.}.
As for communication overhead, we use a straggler-resilient scheme based on backup worker~\cite{Chen2016} to alleviate the overhead caused by heterogeneity. 
For computation overhead, we compress the input matrix based on the principal component analysis~\cite{2014A} and reduce the  computational redundancy caused by homomorphic encryption. 
Our system reduces the communication overhead in heterogeneous scenarios range 13.99\%-65.26\% with different network situations and backup worker settings, and reduces the computation overhead caused by HE range 21.95\%-40.66\% with different degrees of compression. 
Our system can significantly improve the efficiency of the current vertical federated learning without loss of security.
In summary, we have made the following contributions:
\begin{itemize}
    \item We carry out the first VFL system measurement on industrial federated learning framework FATE and demonstrate its main performance bottlenecks.
    \item We propose an accelerating system designed specifically for tackling VFL performance bottlenecks without loss of security.
    \item We demonstrate the effectiveness of our system through extensive experiments. Our system makes VFL more practical for real word settings. 
\end{itemize}

The structure of this paper is as follows. In Section~\ref{background}, we analyze the inspiration of this paper from the perspective of the advantages and bottlenecks of vertical federated learning under homomorphic encryption. Then we propose our system to overcome the existing bottlenecks and accelerate vertical federated learning in Section~\ref{system}. The experimental results showing the efficiency and robustness of our system are shown in Section~\ref{experiment}. Related work is introduced in Section~\ref{related}. Finally, we conclude our paper and put forward several directions for future work in Section~\ref{conclusion}.

%% file: fig-fl.tex
\begin{figure}[!t]
    \hspace{-30pt}
    \centering
    \includegraphics[width=3.3in]{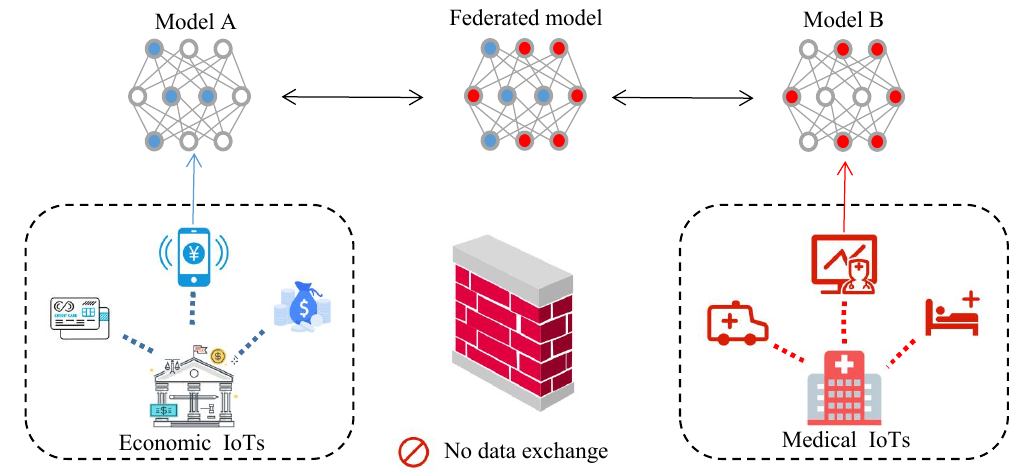}
    \caption{Vertical federated learning. A specialization of federated learning for secure cross-silo cooperation learning.}
    \label{fig-fl}
\end{figure}

%% file: motivation.tex
\section{Background and Motivation}
\label{background}

This part introduces the vertical federated learning framework adopted by FATE~\cite{FATE} and its efficiency in privacy-preservation. 
Then we dive into the bottlenecks of the existing framework and raise our motivation.

\input{architecture.tex}

\input{encryption.tex}

\input{bottleneck.tex}

%% file: architecture.tex
\subsection{Vertical Federated Learning Architecture}

Different from conventional FL, each cooperator has a different feature dimension in vertical federated learning. 
Apart from distributed features, in order to preserve privacy, labels of cooperators will not be shared with others. 
In other words, only the cooperator who seeks collaboration owns the labels. 
We note that the cooperator seeking collaboration and owing labels as Guest. 
And those who attend the collaboration are called Hosts.
The lack of labels and features directly leads to the fact that cooperators can not compute loss locally and thus fail to compute gradients.
So conventional federated learning procedure is not feasible in vertical federated learning.

Hardy et al. propose a secure cooperative learning framework~\cite{Hardy2017} for vertical data using additional homomorphic encryption.
The framework is as precise as a naive non-private solution that brings all data in one place and scales to problems with millions of entities with hundreds of features. 
It consists of two parts. 
The first part is encrypted entity alignment. 
Considering that data samples of different partners are not the same, the system utilizes many encryption allocation schemes to locate these common data samples in multiple parties. 
In this process, the data information that does not coincide with each other will not be disclosed. 
The second part is encrypted model training, which is the focus of our work. 
After encrypted entity alignment, it can be assumed that the data samples of all parties are consistent, but the feature spaces of all parties do not coincide. 
As shown in Fig. \ref{fig-framework}, before the training, the authoritative arbiter randomly generates the private key and public key. 
Then the arbiter distributes the public key to each cooperator for encryption.
Note that cooperators have no right or opportunity to obtain the private key. 
That is, cooperators have only the right to encrypt, but not the right to decrypt, which ensures that the interactive data is presented in ciphertext. 
During one iteration, each cooperator first completes the forward propagation and obtains the intermediate results in plain mode. 
After encrypting the intermediate result with their own public key, cooperators exchange the encrypted intermediate results. 
Then they can gather a complete encrypted loss, containing the data feature information of other parties and label information. 
They use the encrypted loss to compute encrypted gradients and send encrypted gradients to arbiter for decryption. 
After decrypting their respective gradients, the arbiter will return the separated gradient parameters to each cooperator.
\input{fig-framework}

%% file: fig-framework.tex
\begin{figure}[!t]
    \hspace{-30pt}
    \centering
    \includegraphics[width=3.3in]{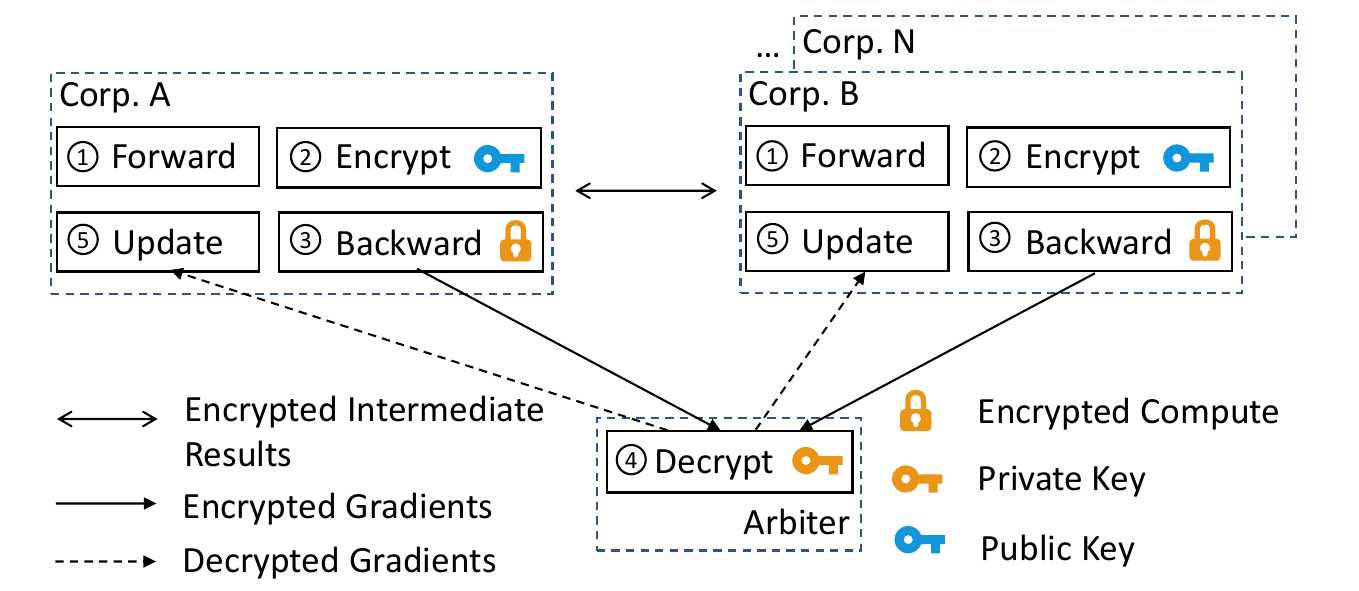}
    \caption{The framework of vertical federated learning, where homomorphic encryption is applied to preserve privacy.}
    \label{fig-framework}
\end{figure}

%% file: encryption.tex
\subsection{Encryption mechanism}
Homomorphic encryption allows direct operations on the ciphertext without decryption, which has attracted a large number of researchers to explore and study in recent years~\cite{Zhang2020, Liu2019, cheng2021secureboost,aono2017privacy}. 
The most commonly used homomorphic encryption algorithm is Paillier~\cite{1999Public}.
In paillier, additive homomorphic is satisfied, which means that plaintext addition is realized through the overloaded operator of ciphertext.
More precisely, the result of ciphertext multiplication after decryption is equivalent to plaintext addition:

\begin{equation}
    D([[ x ]] \bullet [[ y ]])=x+y,
    \label{HE}
\end{equation}

where $[[x]]$ is the ciphertext of x, function $D$ represents decryption.
Similarly, the result of ciphertext exponentiation after decryption is equivalent to plaintext multiplication. 
Note that the size of ciphertext tends to be excessively bigger than the size of plaintext. 
For example, the commonly used plaintext is 8-bit or 16-bit, while the shortest ciphertext is 2048-bit till 2019~\cite{Zhang2020}. 
Homomorphic encryption will greatly increase the complexity of the operation, especially for multiplication.

%% file: bottleneck.tex
\subsection{Performance Bottleneck}
We carry out some preliminary experiments\footnote{Experiments are on \texttt{Epsilon}~\cite{pascal-voc-2008} dataset with 5000 samples and 2000 features separated for four cooperators. 
Computing capacity remains constant during training. 
Training performance of three modes is identical after 50 iterations.} 
to show the performance bottleneck intuitively.
As shown in Fig. \ref{fig-composition}, vertical federated learning mainly consists of four parts, namely \emph{Computation, Encryption, Communication} and \emph{Other} operation. 
\emph{Other} operation is the time of IO operations and task scheduling.  
Since encryption has a negligible impact on IO operations, we can reasonably assume that time consumption of \emph{Other} is static (100s) in all three modes, and use it as a comparing reference. 
Plain mode stands for vertical federated learning without homomorphic encryption and heterogeneity, where \emph{Other} covers the most running time. 
Encrypt mode stands for vertical federated learning under homomorphic encryption without network heterogeneity, which means that no straggler will appear. 
In encrypt mode, \emph{Computation} consumes most of time, \emph{Encryption} consumes second. 
Hetero-encrypt mode is vertical federated learning in practical deployment, where both homomorphic encryption and heterogeneity are involved. 
In hetero-encrypt mode, \emph{Communication} surpasses in reverse. 
Among those three bottlenecks, the increase in encryption time is common, which has been widely studied~\cite{Zhang2020, aono2017privacy}. 
So we dive into those two specific bottlenecks: computation overhead and communication overhead.
\input{fig-composition}
\subsubsection{Computaion Overhead}
Different from the horizontal federated learning, the parameters transmitted in the vertical federated learning are not only gradients, but also intermediate results. 
Moreover, due to the potential competitive relationship, all parties do not have the authority to decrypt the parameters to preserve privacy. 
This means that homomorphic encryption not only brings the redundancy contained in the encryption and decryption computation itself, but also forces all parties to do computation on ciphertext instead of plaintext. 
And the ciphertext computation of homomorphic encryption is very time-consuming, especially for multiplication.

\subsubsection{Communication overhead}
Heterogeneous networks and huge geographical spans are the main reasons for communication overhead. 
More specifically, multiple institutions with different features may be distributed in all corners of the earth, and their geographical positions are generally considerably different. 
Moreover, the network condition of a single organization is difficult to ensure stability due to the network heterogeneity~\cite{Yu2018}. 
Because each institution itself has limited communication resources~\cite{2}, such as carrier, bandwidth, e.t.c. 
These resources may be allocated to other jobs, as cooperators do not exist exclusively for federated learning. 
These services will bring huge network traffic during the peak hours, and even lead to network congestion.
What is worse, homomorphic encryption will expand the data size, which amplifies the impact of network fluctuation.

The above two bottlenecks cause much idle time in vertical federated learning. 
That is, one party may spend more time waiting for the calculation results than other parties. 
Such random and unbalanced idle time brings difficulty to the application and promotion of vertical federated learning.

%% file: fig-composition.tex
\begin{figure}[!t]
    \hspace{-30pt}
    \centering
    \includegraphics[width=3.3in]{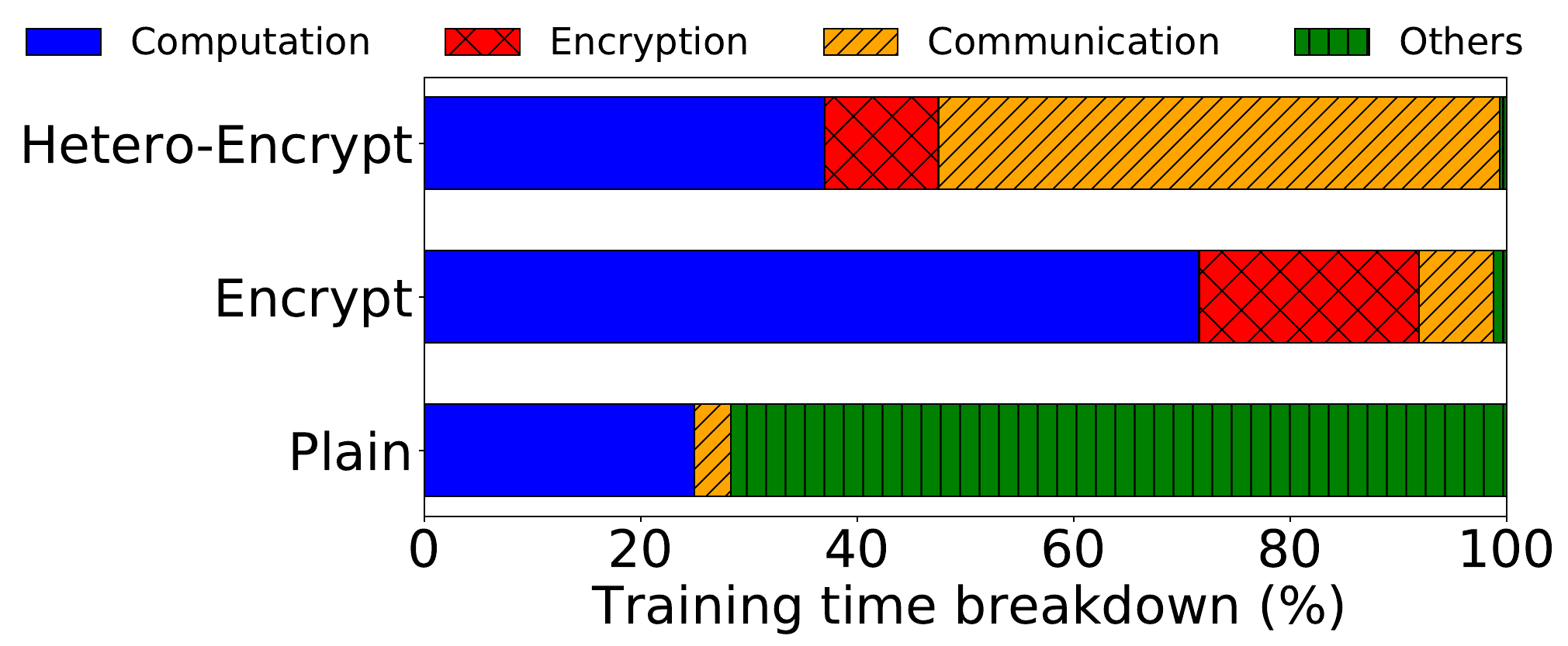}
    \caption{Runtime breakdown under different VFL settings. }
    \label{fig-composition}
\end{figure}

%% file: method.tex
\section{Proposed System}
\label{system}

In order to reduce the communication and computation overhead in vertical federated learning caused by homomorphic encryption and heterogeneity, we propose a practical and effective optimization system under the industrial federated learning framework FATE~\cite{FATE}.

\input{definition}

\input{method-backup.tex}

\input{method-pca.tex}


%% file: definition.tex
\subsection{Problem Definition}
\label{problem}
We suppose one cooperator seeks help for $K$ cross-industry cooperators. 
For simplicity, we call cross-industry cooperators without labels as Hosts, denoted as $A_{k}$, for $k = 1,2..K$. Similarly, we call the cooperator seeking help and owing labels as Guest, denoted as $B$. Assuming learning rate $\mu$, regularization parameter $\lambda$, 
dataset set $\left\{x_{i}^{A_{k}}\right\}_{i \in \mathcal{D}_{A_{k}}},\left\{x_{i}^{B}, y_{i}\right\}_{i \in \mathcal{D}_{B}}$, and model 
paramters $\Theta_{A_{k}}, \Theta_{B}$ corresponding to the feature space of $x_{i}^{A_{k}}$,  $x_{i}^{B}$ respectively, the training
objective is:

\begin{equation}
    \begin{aligned}
    \min _{\Theta_{A_{k}}, \Theta_{B}} 
    &\sum_{i}\left\|\sum_{k=1}^{K}\Theta_{A_{k}} x_{i}^{A_{k}} +\Theta_{B} x_{i}^{B}-y_{i}\right\|^{2}\\
    &+\frac{\lambda}{2}\left(\sum_{k=1}^{K}\left\|\Theta_{A_{k}}\right\|^{2}+\left\|\Theta_{B}\right\|^{2}\right).
    \label{algorithm-loss}
    \end{aligned}
\end{equation}

Let $u_{i}^{A_{k}}=\Theta_{A_{k}} x_{i}^{A}, u_{i}^{B}=\Theta_{B} x_{i}^{B}$, the encrypted loss is:

\begin{equation}
    \begin{aligned}
    \left[\left[\mathcal{L}\right]\right]=[[\sum_{i}\left\|\sum_{k=1}^{K}u_{i}^{A_{k}}+u_{i}^{B}-y_{i}\right\|^{2} \\
    +\frac{\lambda}{2}\left(\sum_{k=1}^{K}\|\Theta_{A_{k}}\|^{2}+\|\Theta_{B}\|^{2}\right)]],
    \label{algorithm-encrypted-loss}
    \end{aligned}
\end{equation}

where additive homomorphic encryption is denoted as $[[\cdot]]$. Let $[[\mathcal{L}_{A}]]=[[\sum_{i}\left(\sum_{k=1}^{K}u_{i}^{A_{k}}\right)^{2}+\frac{\lambda}{2} \sum_{k=1}^{K}\Theta_{A_{k}}^{2}]]$,
$[[\mathcal{L}_{B}]]=[[\sum_{i}\left(\left(u_{i}^{B}-y_{i}\right)^{2}\right)+\frac{\lambda}{2} \Theta_{B}^{2}]]$, 
and $[[\mathcal{L}_{A B}]]=2 \sum_{i}\sum_{k=1}^{K}\left([[u_{i}^{A_{k}}]]\left(u_{i}^{B}-y_{i}\right)\right)$, then
\begin{equation}
    [[\mathcal{L}]]=[[\mathcal{L}_{A}]]+[[\mathcal{L}_{B}]]+[[\mathcal{L}_{A B}]].
    \label{algorithm-L}
\end{equation}

Note that an additively homomorphic encryption scheme provides an operation that produces the sum of two numbers under encryption.
For simplicity, we overload the notation and we denote the operator with "$+$" as well. Similarly, let $[[d_{i}]]=\sum_{k=1}^{K}[[u_{i}^{A_{k}}]]+[[u_{i}^{B}-y_{i}]]$, then gradients are:

\begin{equation}
    \begin{aligned}
    \left[\left[\frac{\partial \mathcal{L}}{\partial \Theta_{A_{k}}}\right]\right]
    &=\sum_{i}[[d_{i}]] x_{i}^{A_{k}}+[[\lambda \Theta_{A_{k}}]], \\
    \left[\left[\frac{\partial \mathcal{L}}{\partial \Theta_{B}}\right]\right]
    &=\sum_{i}[[d_{i}]] x_{i}^{B}+[[\lambda \Theta_{B}]].
    \end{aligned}
    \label{algorithm-gradient}
\end{equation}

So, for iteration $j=1,2...$, updated model parameters are:

\begin{equation}
    \begin{aligned}
        \Theta_{A_{k}}^{j+1}
        &=\Theta_{A_{k}}^{j}-\mu\frac{\partial \mathcal{L}}{\partial \Theta_{A_{k}}}, \\
        \Theta_{B}^{j+1}
        &=\Theta_{B}^{j}-\mu\frac{\partial \mathcal{L}}{\partial \Theta_{B}}.
    \label{algorithm-update}
    \end{aligned}
\end{equation}

%% file: method-backup.tex
\subsection{Backup Worker}
Cooperators in vertical federated learning need to exchange encrypted intermediate results. 
As can be seen from the formula (\ref{algorithm-gradient}), each cooperator can calculate the gradients with local data after 
getting the global $d$. 
The critical point of vertical federated learning is to calculate the global $d$ cooperatively. 
As mentioned in Section \ref{background}, 
the network environment is likely to be heterogeneous and volatile. The existing synchronization framework 
will lead to much idle time, which will damage the efficiency and robustness of the training process.

The simple idea is to discard the unreachable parameters, which is effective in distributed learning because only gradients are transmitted from partial workers~\cite{Chen2016}. 
But considering the difference between VFL and conventional distributed learning, such an arbitrary drop scheme may not perform well in VFL. As shown in Fig. \ref{fig-framework}, cooperators in VFL exchange gradients with arbiter to decrypt gradient instead of aggregating gradients.
This means that dropping gradients can block the learning progress or degrade the clients to training locally without collaboration. 
Variables transmitted between cooperators are encrypted intermediate results, which will be used for gradient computation.
So arbitrary discard will inevitably damage training performance. The disadvantage of this method is magnified in the longitudinal federated learning, because some clients will accumulate such damages and turn into performance stragglers, which means its misleading drags the global performance.

Therefore, in order to enhance the stability of communication without excessive loss of accuracy, we combine the Stale Synchronous Parallel Parameter Server (SSP)~\cite{7} scheme with the original backup scheme~\cite{Chen2016}, that is, using the old backup data to fill in the parameters corresponding 
to the missing positions in this round. The detailed algorithm is shown in Algorithm \ref{algo-1}. 
Assuming $K$ hosts with dataset $x_{i}^{A_{k}}$ and one guest with dataset $x_{i}^{B}$ attend the cooperation, we set the number of backup workers as $\beta$.
Hosts compute $u_{i}^{A_{k}}$ with local dataset, and encrypt it with homomorphic encryption for secure exchange. Guest waits for $u_{i}^{A_{k}}$ from $A_{k}$. After receiving the first $K-\beta$ parameters in each iteration, the guest stops waiting for data from other parties, and then uses the stale backup of the unreceived party to compute $[[\mathcal{L}]]$ and $[[d_{i}]]$.
\begin{equation}
    \begin{aligned}
        \left[\left[\mathcal{L}\right]\right]=[[\sum_{i}\left\|\sum_{s=1}^{t}u_{i}^{A_{j_{s}}}+\sum_{s=t+1}^{K}\hat{u_{i}^{A_{j_{s}}}}+u_{i}^{B}-y_{i}\right\|^{2} \\
    +\frac{\lambda}{2}\left(\sum_{s=1}^{t}\|\Theta_{A_{j_{s}}}\|^{2}+\sum_{s=t+1}^{K}\|\hat{\Theta_{A_{j_{s}}}}\|^{2}+\|\Theta_{B}\|^{2}\right)]],
    \label{algorithm-backup}
    \end{aligned}
\end{equation}

\begin{equation}
[[d_{i}]]=\sum_{s=1}^{t}[[u_{i}^{A_{j_{s}}}]]+\sum_{s=t}^{K}[[\hat{u_{i}^{A_{j_{s}}}}]]+[[u_{i}^{B}-y_{i}]],
\label{algorithm-backup2}
\end{equation}

where $\hat{u_{i}^{A_{j_{s}}}}, \hat{\Theta_{A_{j_{s}}}}$ is the $j_{th}$ iteration's backup variables, $t=K-\beta$, $j_{s}$ is the receieve sequence of $j_{th}$ iteration. 
The backup scheme can be various, such as average, local prediction and so on. For the convenience of discussion, the simplest method is chosen, that is, to fill the unreceived parameters with the results in the previous iteration. The experimental part below shows that even this simple scheme has better effects than the arbitrary discard scheme.

\input{algorithm-backup.tex}

\textbf{Remark.}
Backup worker allows distributed workers to use older, stale versions of model's values (gradients and logits).
This significantly helps workers to spend more time on computing instead of idle waiting.
Furthermore, the SSP model~\cite{7} can ensures the correctness of ML algorithm by limiting the maximum age of the stale values.
Considering a nearly half proportion of encrypted computation in VFL training (shown in Fig. \ref{fig-composition}), common network fluctuation (the best network bandwidth is within one magnitude of the worst) will not incur the age of stale value $\geqq$ 2. 
In this circumstance, the SSP model can perform on par with vanilla synchronous training.
Note that our system does not use any intelligence client selection mechanism~\cite{nishio2019client, xu2020client, wang2021device}, but is compatible with above techniques.
These techniques can be injected to further ensure the upper bound of the stale age.

%% file: algorithm-backup.tex
\begin{algorithm}
    \caption{Backup for VFL}
    \KwIn{dataset $x_{i}^{A_{k}}$, dataset $x_{i}^{B}$, backups $\beta$  }
    \KwOut{Model parameters $\Theta_{A_{k}}, \Theta_{B}(k=1,2..K)$}
    Party $A_{k},B$ initialize $\Theta_{A_{k}}, \Theta_{B}$;
    
    \For{each iteration $j=1,2 \ldots$}{
        Parallel: Host(j), Guest(j)
    }
    Host(j):

    \For{k=1,2..K in parallel}{
        $A_{k}$ computes $u_{i}^{A_{k}}$, and send $[[u_{i}^{A_{k}}]], [[\Theta_{A_{k}}^{2}]]$ to B

        Get $[[d_{i}]]$ from $B$

        Updata $\Theta_{A_{k}}$ with (\ref{algorithm-gradient}) (\ref{algorithm-update})
    }

    Guest(j):

    \eIf{$\left|[[u_{i}^{A_{k}}]]\right|< K-\beta$}{
        Waiting for $u_{i}^{A_{k}}$ from $A_{k}$
    }{
        B computes $d_{i}$ with (\ref{algorithm-backup2}) and sends $[[d_{i}]]$ to $A_{k}$
    }
    Updata $\Theta_{B}$ with (\ref{algorithm-gradient}) (\ref{algorithm-update})
    \label{algo-1}
\end{algorithm}

%% file: method-pca.tex
\subsection{Dynamic Feature Selection}
Multiplication of homomorphic encryption consumes a huge computation cost. 
We assume that the host data set $x_{i}^{A_{k}}$ has $m$ data samples and $n$ features, noted as $X_{m \times n}$.
So a gradient calculation $[[d_{i}]] x_{i}^{A_{k}}$ in (\ref{algorithm-gradient}) requires $m \times n$ times of encrypted multiplication. This leads to the huge computation overhead in vertical federated learning, the experimental results have been shown in Fig. \ref{fig-composition}.

We compress the feature quantity by the dimension reduction method called Principal Component Analysis (PCA)~\cite{2014A} to reduce the frequency of multiplication, shown in (\ref{pca-1}). 
\begin{equation}
    Z_{m \times k}=f\left(X_{m \times n}\right), k<n,
    \label{pca-1}
\end{equation}

where $Z_{m \times k}$ is the compressed input matrix, $f$ is a mapping matrix to maximize the variance of data under the specified dimension $K$, note $f$ as $W_{k \times n}^{T}$.

The traditional principal component analysis is only a one-time dimensionality reduction, which can not be 
organically combined with current vertical federated learning. Therefore, we design a scheme shown in Algorithm \ref{algo-2} so 
that principal component analysis can be effectively integrated into vertical federated learning. At the beginning of each iteration, cooperators generate a compression matrix $W_{k \times n}^{T}$, the target dimension $k$ can be set flexibly to find a balance in efficiency and accuracy. First, cooperators compress the input matrix $\Theta_{1 \times n}, X_{n \times m}$ into $\theta_{1 \times k}, Z_{k \times m}$.

\begin{equation}
    \begin{aligned}
    \theta_{1 \times k} = \Theta_{1 \times n} W_{k \times n}^{T}, \\
    Z_{k \times m}=W_{k \times n} X_{n \times m}.
    \label{pca-2}
\end{aligned}
\end{equation}

Then they can multiply the compressed input matrix with encrypt $[[d_{i}]]$ using (\ref{algorithm-L}) (\ref{algorithm-gradient}) to get gradients $[[\frac{\partial \mathcal{L}}{\partial \Theta_{A_{k}}}]]$ with $k$ gradients, noted as $g_{1 \times k}$. The compression mechanism turns the encrypted matrix from $X_{n \times m}$ to $Z_{k \times m}$ and thus reduces 
$m \times n$ times of homomorphic encryption multiplication to $m \times k$ times.

At the end of each iteration, cooperators will decompress the decrypted gradient. 
\begin{equation}
    \begin{aligned}
    g_{1 \times n} = g_{1 \times k} W_{k \times n}^{-1}.
    \label{pca-3}
    \end{aligned}
\end{equation}

\textbf{Remark} Compression and decompression are both performed on local datasets in plaintext. 
The computation overhead of plaintext is nearly negligible compared to that of ciphertext multiplication. 
And our compression behavior is recoverable by storing the PCA matrix locally.
It avoids the inexplicability caused by compressing features on the air, and ensures that the algorithm can perfectly fit the existing VFL framework without enormous modification.
\input{algorithm-pca.tex}

%% file: algorithm-pca.tex
\begin{algorithm}
    \caption{PCA for VFL}
    \KwIn{dataset $X_{m \times n}$  }
    \KwOut{Model parameters $\Theta_{1 \times n}$}
    Initialize $\Theta_{1 \times n}$;
    
    \For{each iteration $j=1,2 \ldots$}{
        Generate the PCA matrix $W_{k \times n}^{j}$

        Compress the input matrix with (\ref{pca-2})

        VFL Train with (\ref{algorithm-L}) (\ref{algorithm-gradient})

        Unzip gradient $g_{1 \times k}$ with (\ref{pca-3})
        
        Update $\Theta_{1 \times n}^{j}$ with (\ref{algorithm-update})
    }
    \label{algo-2}
\end{algorithm}

%% file: experiments.tex
\section{Experiments}
\label{experiment}

We carry out experiments on the industrial federated learning framework FATE~\cite{FATE}. 
Our accelerating system dramatically improves the efficiency and robustness of the current vertical federated learning without loss of security. 

\subsection{Setting}
\textbf{Federated Learning Platform:} 
We build our system on FATE~\cite{FATE}.
FATE is the most popular federated learning framework in Linux and has built cooperation with many enterprises in 
many fields, such as industry, finance, medical, etc.

\textbf{Datasets:} We use three datasets \texttt{MIMIC-III}~\cite{0MIMIC}, \texttt{Epsilon}~\cite{pascal-voc-2008} and \texttt{NUS-WIDE}~\cite{chua2009nus}. 
\texttt{MIMIC-III} is a real-time medical database comprising information about patients admitted to critical care units at a large tertiary care hospital, including vital signs, medications, survival data and more. 
Following the data processing procedures of~\cite{0Multitask}, we get the final training and testing sets of 14681 data samples and 714 features. 
Then we shuffle it into 4 datasets with 114, 200, 200, 200 features.
Datasets are dispatched to 4 cooperators: Guest, Host$_{1}$, Host$_{2}$ and Host$_{3}$ respectively.
\texttt{Epsilon} dataset is instance-wisely scaled to unit length and feature-wisely normalized to mean zero and variance one. 
The raw dimension of \texttt{Epsilon} is 5000 data samples and 2000 data features.
We shuffle it into 4 datasets with 200, 600, 600, 600 features and dispatch the shuffled datasets to guest and hosts respectively.
The \texttt{NUS-WIDE} dataset consists of 634 low-level images features extracted from Flickr images~\cite{mislove2008growth} as well as their associated tags and ground truth labels. 
We put 34 low-level image features on party Guest and 200 features on three Host parties respectively
\footnote{Our micro experiments show that unbalanced features have the almost consistent runtime performance as balanced features. 
Because in the forward propagation, the computation overhead caused by redundant features is in plaintext, which has negligible impact on runtime performance. 
Hence, we reasonably assume that hosts' features are balanced.}.

\textbf{Models:} We use Logist Regression, a common specialization of Linear Regression, to verify the validity of our system. 
We set the best hyper-parameter at our best considering the efficiency and training performance. 
Default hyper-parameters are as follows:
optimizer is rmsprop; learning rate is 0.05; batchsize for \texttt{MIMIC-III} and \texttt{NUS-WIDE} is 1024, and 5000 for \texttt{Epsilon}.

\input{fig-backup-speedup.tex}
\subsection{Backup Worker}

Our system is robust under various network environments and can reduce considerable communication overhead. 
We evaluate our system performance under two different heterogeneous network environments with a baseline bandwidth of 10Mb.
The degree of heterogeneity is quantified by the possibility of slow down, i.e., reaching the bottom bandwidth (1/10 of baseline bandwidth) during network fluctuation~\cite{xu2020client}.
There are three Hosts and one Guest attending the cooperation training. 
As shown in Fig. \ref{fig-backup-speedup}, we use a stable bandwidth network as a baseline, called \emph{Clean}. 
In the situation of 1/4 slow down, the bandwidth will be reduced to 1Mb in the possibility of 1/4. 
$B_{0}$ represents the number of backup workers is 0, which equals the vanilla vertical federated learning in a heterogeneous network. 
The communication overhead is 2.86 times that of \emph{Clean} situation. 
We set the number of backup workers to 1, 2, it will reduce the communication overhead by 13.99\% and 34.97\% respectively.
When the probability of network fluctuation rises to 1/2, which means that the network condition becomes more unstable, network heterogeneity will lead to a sharp increase in network communication overhead. 
Vertical federated learning under 1/2 slow down without optimization ($B_{0}$) will be $7.34 \times$ slower than \emph{Clean}.
The effect of our system will be more obvious. When the number of backups is 1, 2 respectively, the communication overhead can be reduced by 47.41\% and 65.26\% respectively.

\input{fig-backup-performance.tex}

And this dramatic reduction in communication overhead does not affect the training performance.
As shown in Fig. \ref{fig-backup-performance}, when the number of backup workers is set to 1, the training process and final performance are almost the same as vanilla VFL ($B_{0}$).
When the number of backup workers is set to 2, the final training performance is almost the same as $B_{0}$, but the training process will become unstable. 
In Fig. \ref{fig-backup-staleness}, we compare the effect of two different backup schemes.
The conventional backup scheme ignoring the unreceived parameters has a noticeable loss of convergence performance in three datasets.
This is because some critical information of the global training loss is dropped, leading to a huge deviation when computing encrypted gradients. 
In contrast, compensating the unreceived parameters with the stale variants can effectively narrow the deviation when the staleness age is limited~\cite{7} and perform on par with vanilla VFL.

\subsection{Principle Component Analysis}

As shown in Fig. \ref{fig-pca-performance-step}, the feature compression method called principal component analysis has different performance on different datasets. Origin means vanilla VFL without optimization. 
PCA converts the input dimension to the target compressed dimension.
In Fig. \ref{fig-pca-performance-step} (a) and (c), our feature compression scheme that reduces the feature dimension to 60\% does not cause a significant loss of accuracy on \texttt{MIMIC-III} and \texttt{NUS-WIDE}. 
It even converges faster due to the pre-analysis brought from PCA. 
While on the \texttt{Epsilon} dataset shown in Fig. \ref{fig-pca-performance-step} (b), training performance under PCA compression leads initially but fails a little when convergence. 
In other word, although the data after the initial dimensionality reduction using principal component analysis performs better than the original method, the final training performance still suffers a certain degree of accuracy loss. 
With the increase of compression degree, the loss of accuracy will become more apparent. 
Reducing the feature dimension to 60\% on \texttt{Epsilon} will lose nearly 3\% precision when convergence. 
Such a `dataset affinity' may come from the limitations of the principal component analysis, which compressed the feature dimension before training. 
The compression degree of which should be cautiously decided according to the intrinsic correlation of datasets. 
\input{fig-pca-performance-step}
\input{fig-pca-performance-time.tex}

Nonetheless, the possible accuracy loss cannot conceal the enormous increase of runtime performance brought by feature compression. 
As shown in Fig. \ref{fig-pca-performance-time}, after compressing the input dimension via PCA, runtime performance is significantly improved for all three datasets. 
For example, compressing the features dimension of dataset \texttt{MIMIC-III} to 80\% and 60\%, the convergence speed is increased by 21.95\% and 40.66\% compared with the original scheme.
Analogously, training performance of \texttt{Epsilon} and \texttt{NUS-WIDE} is significantly improved as well.
This performance speedup comes from the linearly reduction of encrypted computation, which accounts for around half of the VFL training time.








\subsection{Mixture}
\input{tab-performance.tex}
The above two methods are orthogonal and can be used in combination. 
The global efficiency and robustness of the system can be enhanced by mutual compensation, which means when backup scheme detects stragglers, we can set a higher compression degree to speed up their computation.
As shown in TABLE \ref{tab-mixture}, the backup scheme reduces the communication delay by up to 65\%, the PCA scheme reduces the calculation delay by up to 
40.6\%. Idle time caused by communication and computation overhead is reduced by 55.6\% using two schemes jointly, which will significantly accelerate current vertical federated learning.




%% file: fig-backup-speedup.tex
\begin{figure}[!t]
    \centering
    \includegraphics[width=2.5in]{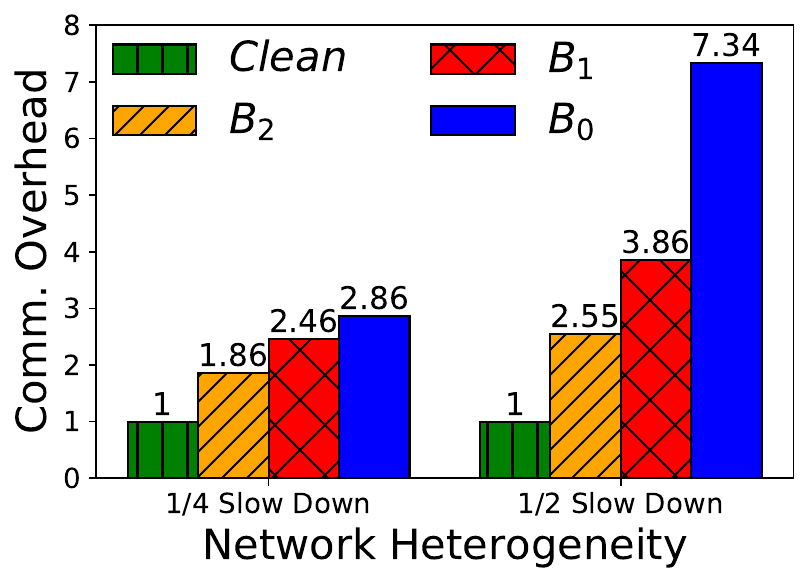}
    \caption{Effect of backup workers: communication speedup. $Clean$ means no heterogeneity, $B_{i}$ means the number of backup workers is $i$ in various heterogeneity (different possibilities of slow down). Dataset: \texttt{MIMIC-III}.}
    \label{fig-backup-speedup}
\end{figure}

%% file: fig-backup-performance.tex
\begin{figure}[ht]			
	\begin{minipage}[b]{0.15\textwidth}
	\includegraphics[width=\textwidth]{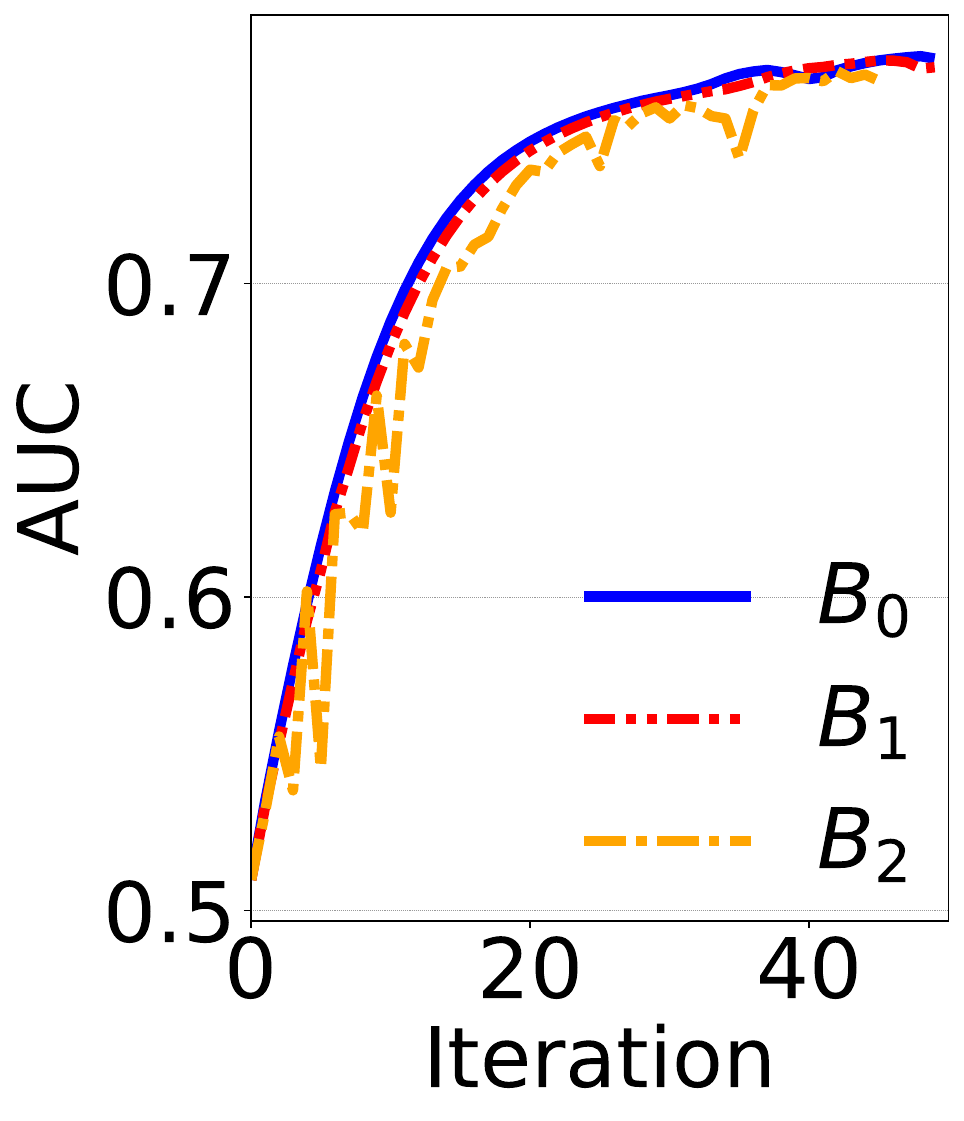}
	\subcaption{MIMIC}
	\end{minipage} 
    ~
	\begin{minipage}[b]{0.15\textwidth}
	\includegraphics[width=\textwidth]{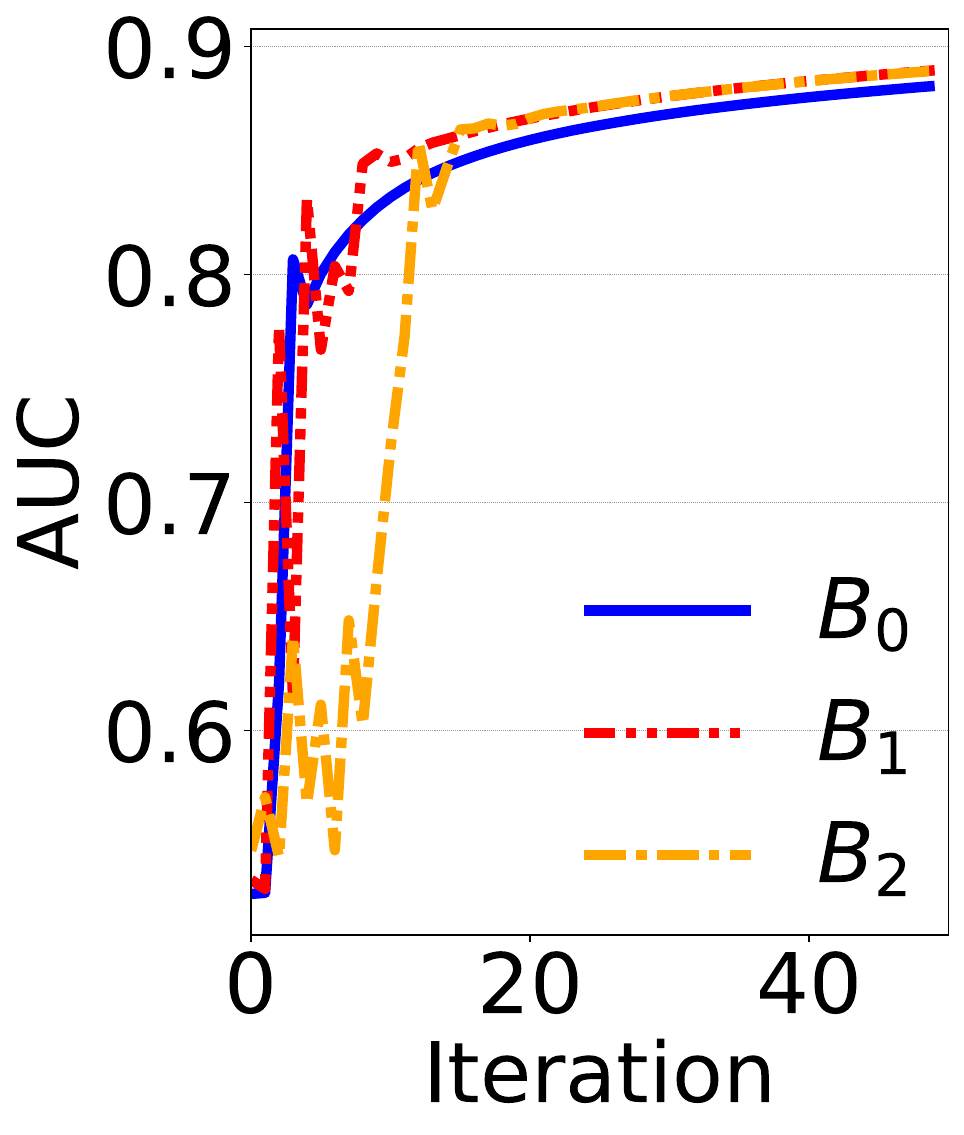}
	\subcaption{Epsilon}
	\end{minipage}
	~
	\begin{minipage}[b]{0.15\textwidth}
	\includegraphics[width=\textwidth]{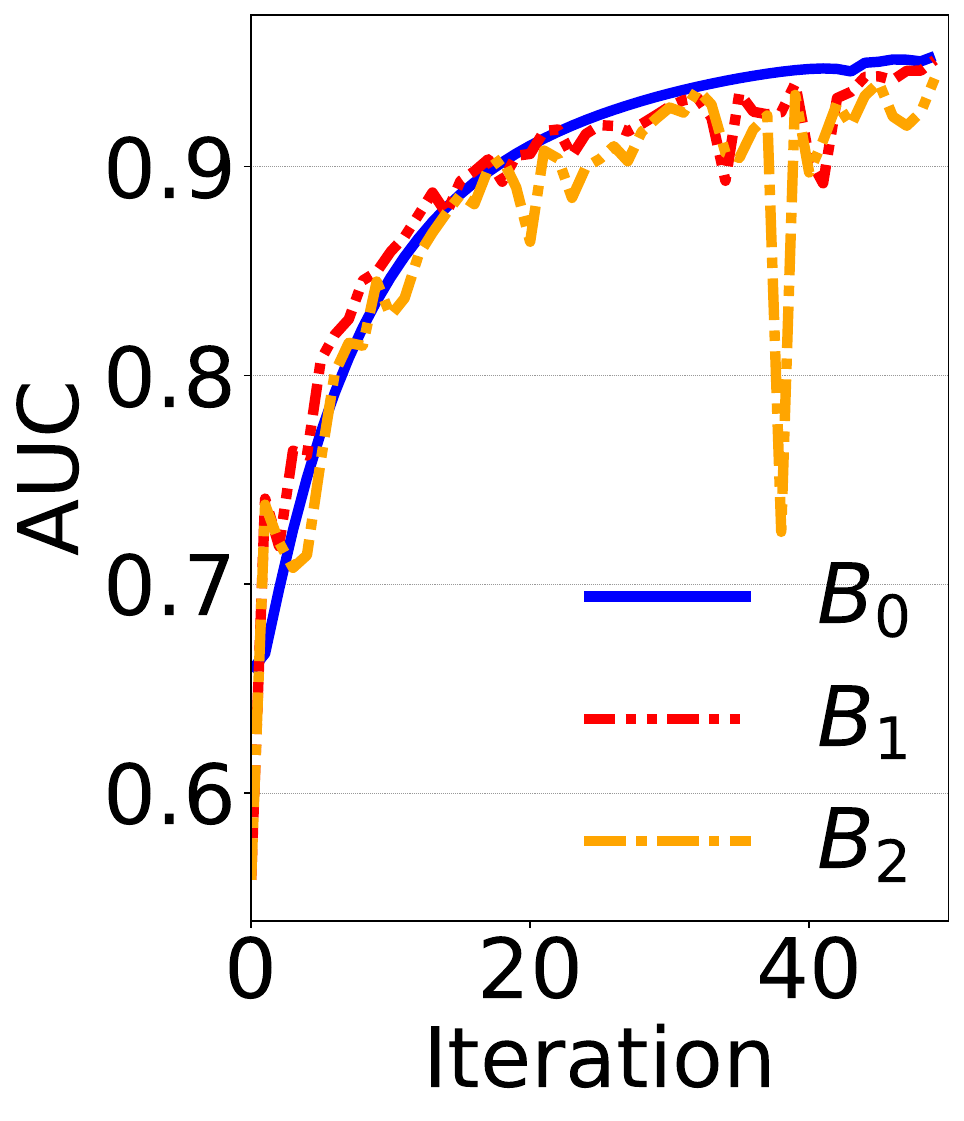}
	\subcaption{NUS-WIDE}
	\end{minipage}
	\caption{Effect of backup workers: AUC vs Iteration.}
	\label{fig-backup-performance}
	
\end{figure}

\begin{figure}[ht]			
	\begin{minipage}[b]{0.15\textwidth}
	\includegraphics[width=\textwidth]{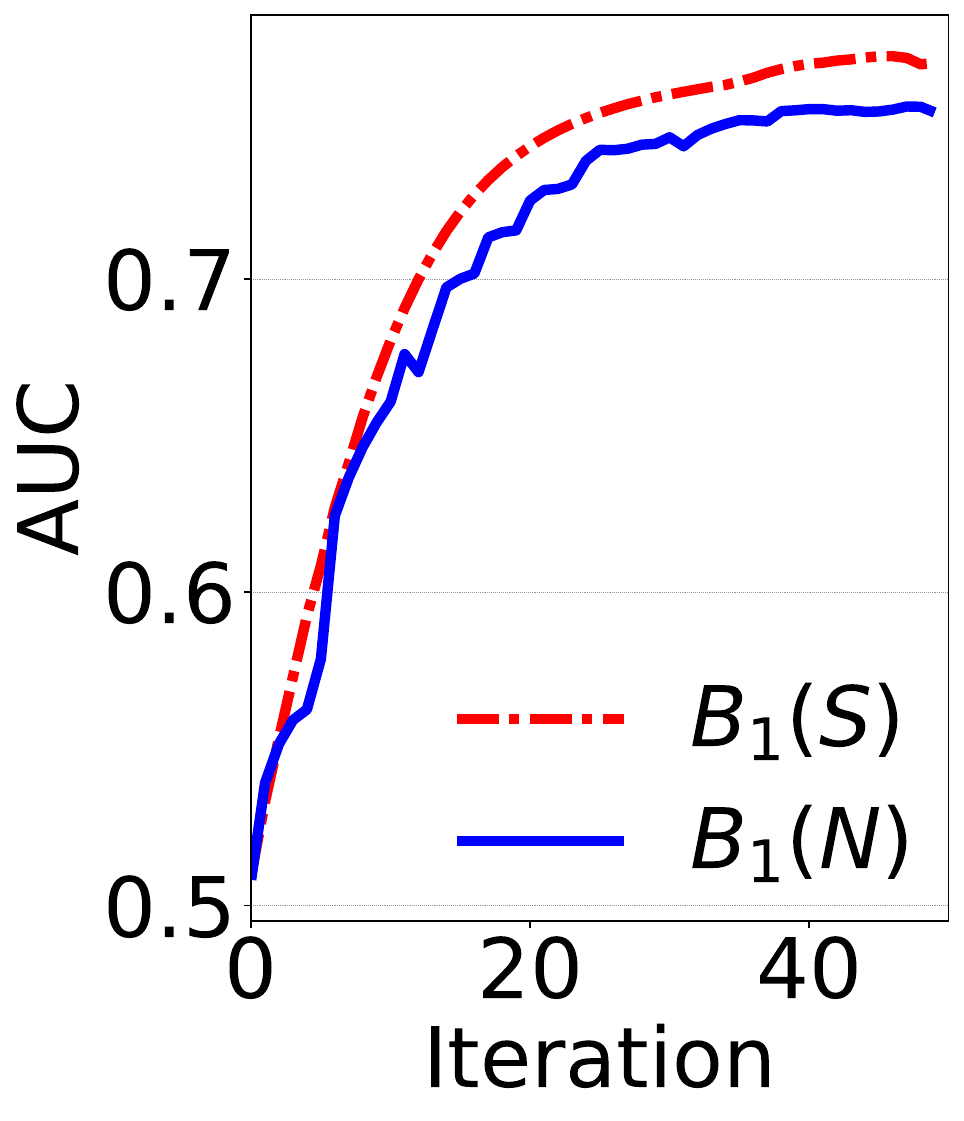}
	\subcaption{MIMIC}
	\end{minipage} 
    ~
	\begin{minipage}[b]{0.15\textwidth}
	\includegraphics[width=\textwidth]{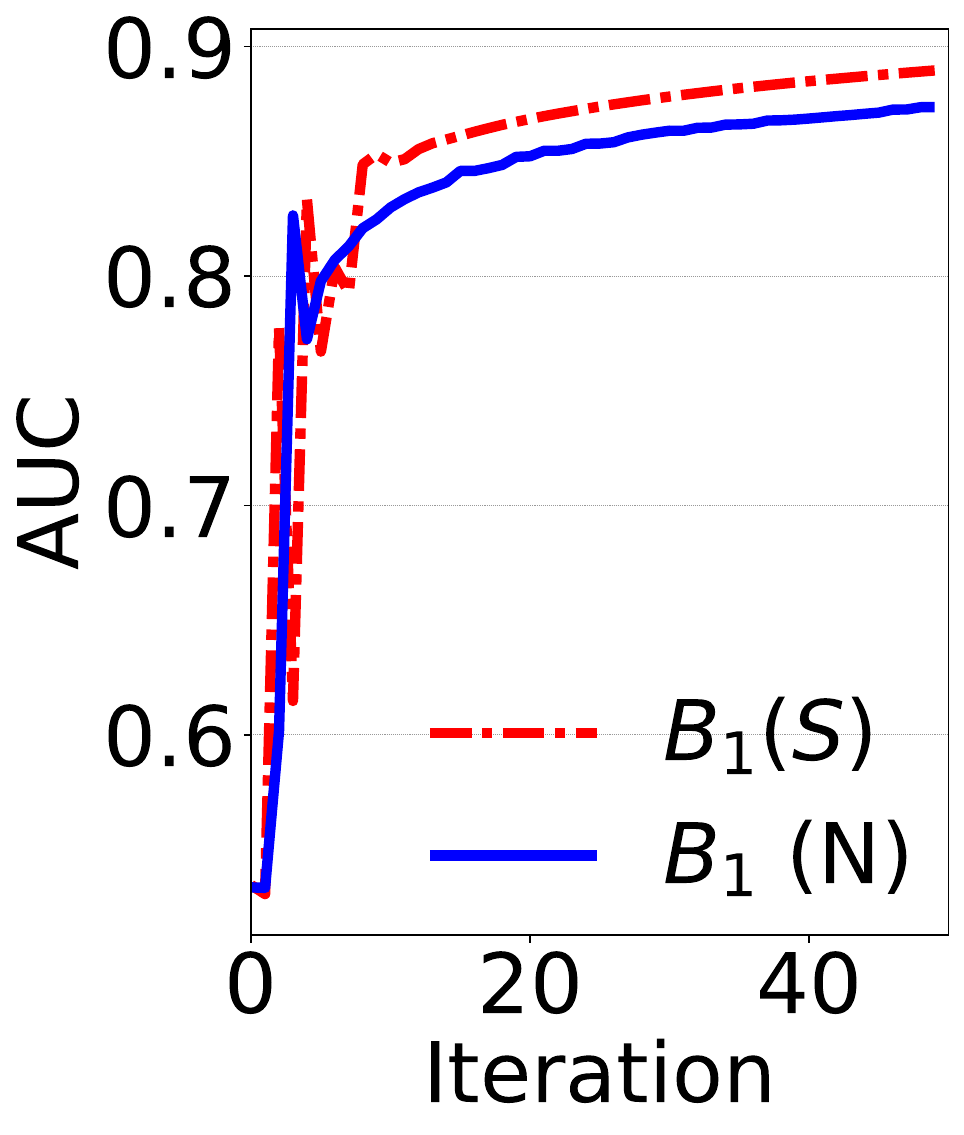}
	\subcaption{Epsilon}
	\end{minipage}
	~
	\begin{minipage}[b]{0.15\textwidth}
	\includegraphics[width=\textwidth]{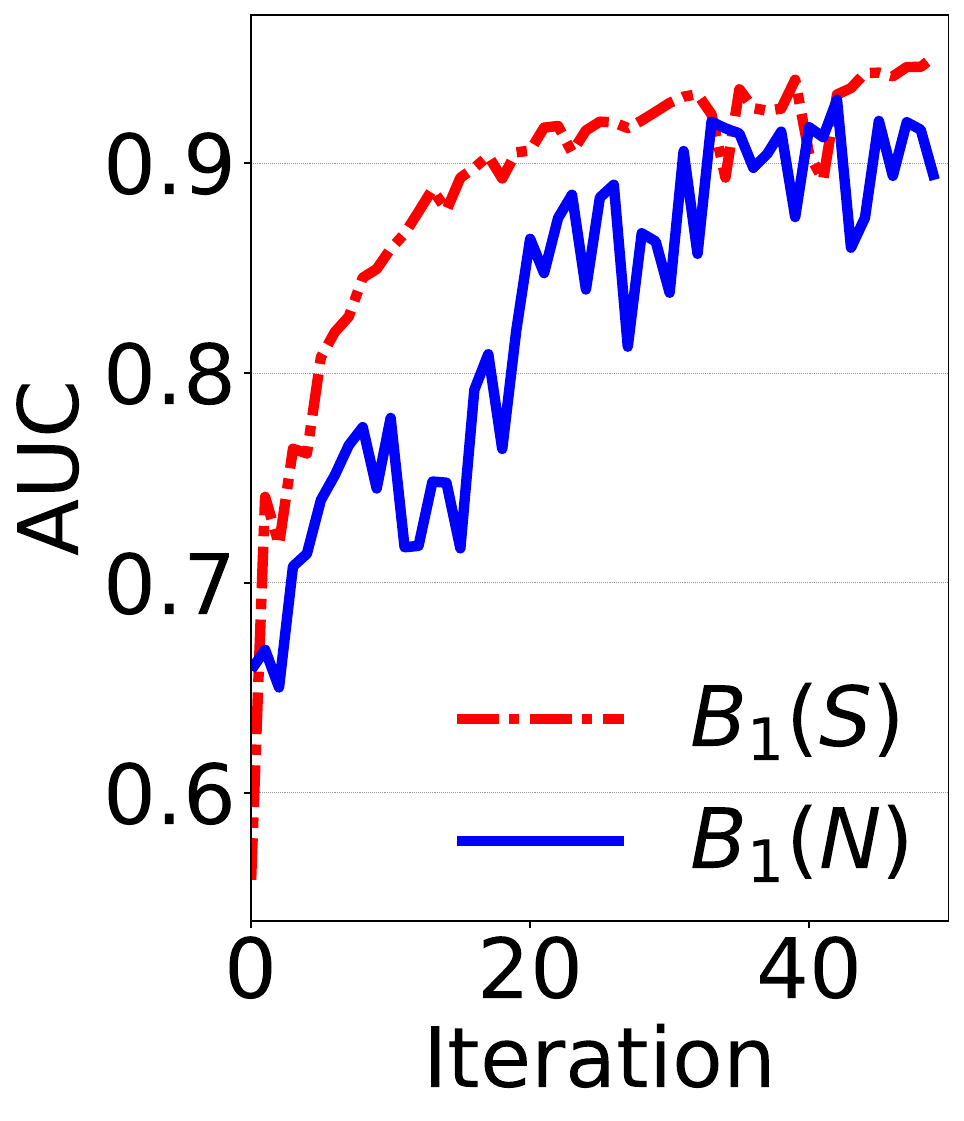}
	\subcaption{NUS-WIDE}
	\end{minipage}
	\caption{Effect of different backup schemes. \textbf{S}tale scheme means to compensate the unreceived parameters with the stale variant, \textbf{N}ull scheme means unreceived parameters are arbitary dropped.}
	\label{fig-backup-staleness}
	
\end{figure}

%% file: fig-pca-performance-step.tex
\begin{figure}[t]			
	\begin{minipage}[b]{0.145\textwidth}
		\includegraphics[width=\textwidth]{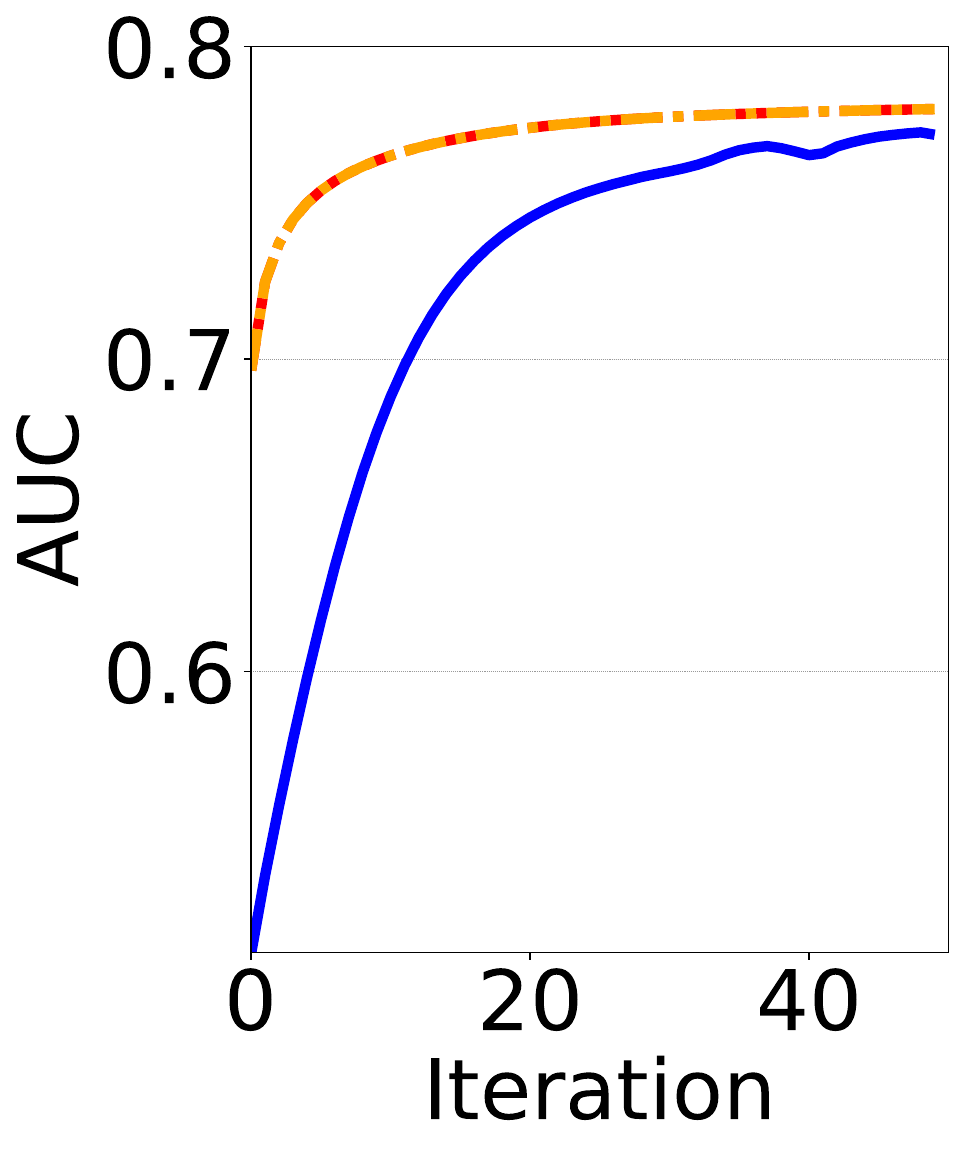}
		\subcaption{MIMIC}
	\end{minipage}
    ~
	\begin{minipage}[b]{0.15\textwidth}
		\hspace*{-35pt}\includegraphics[width=2.12\textwidth]{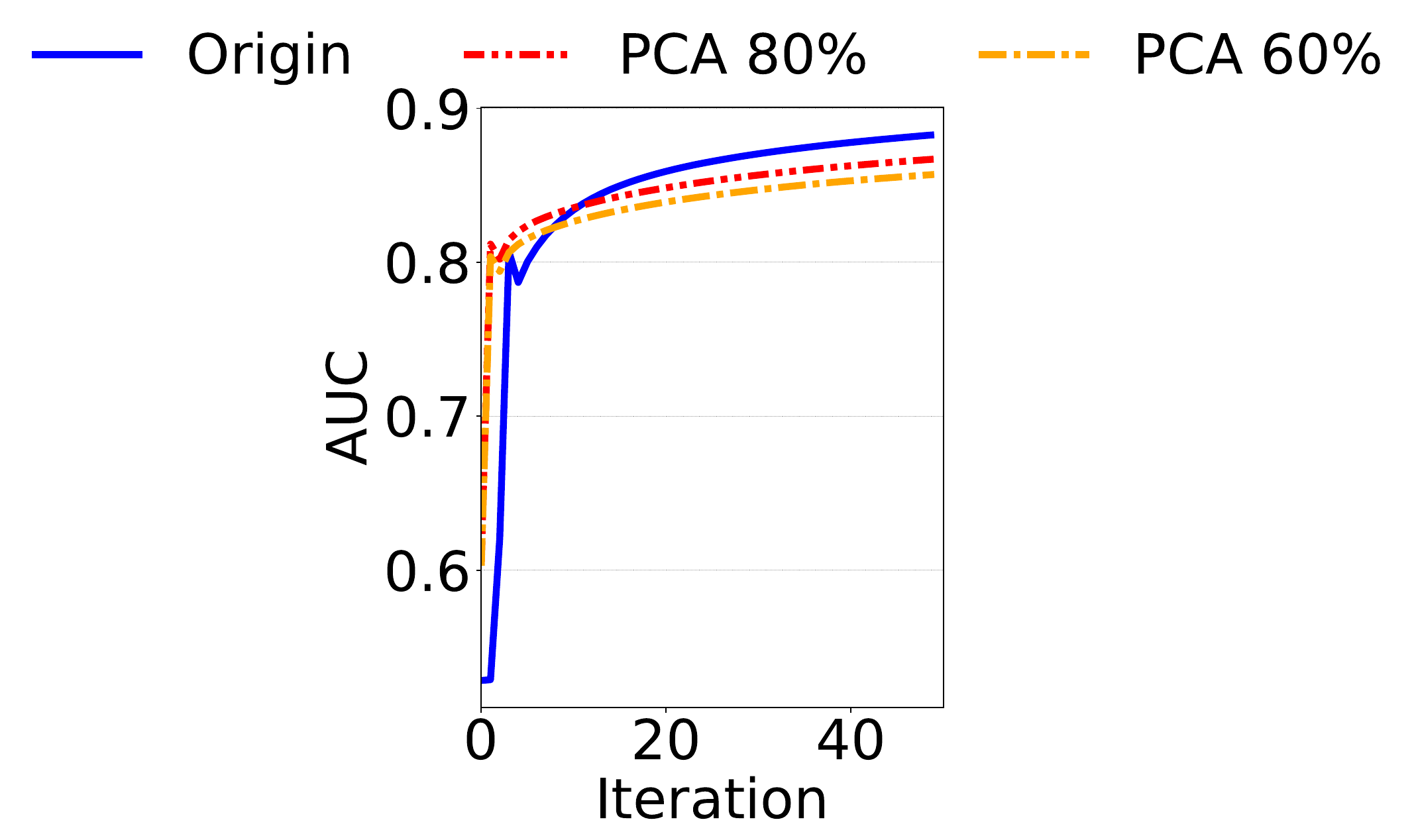}
		\subcaption{Epsilon}
	\end{minipage} 
	~
	\begin{minipage}[b]{0.145\textwidth}
		\includegraphics[width=\textwidth]{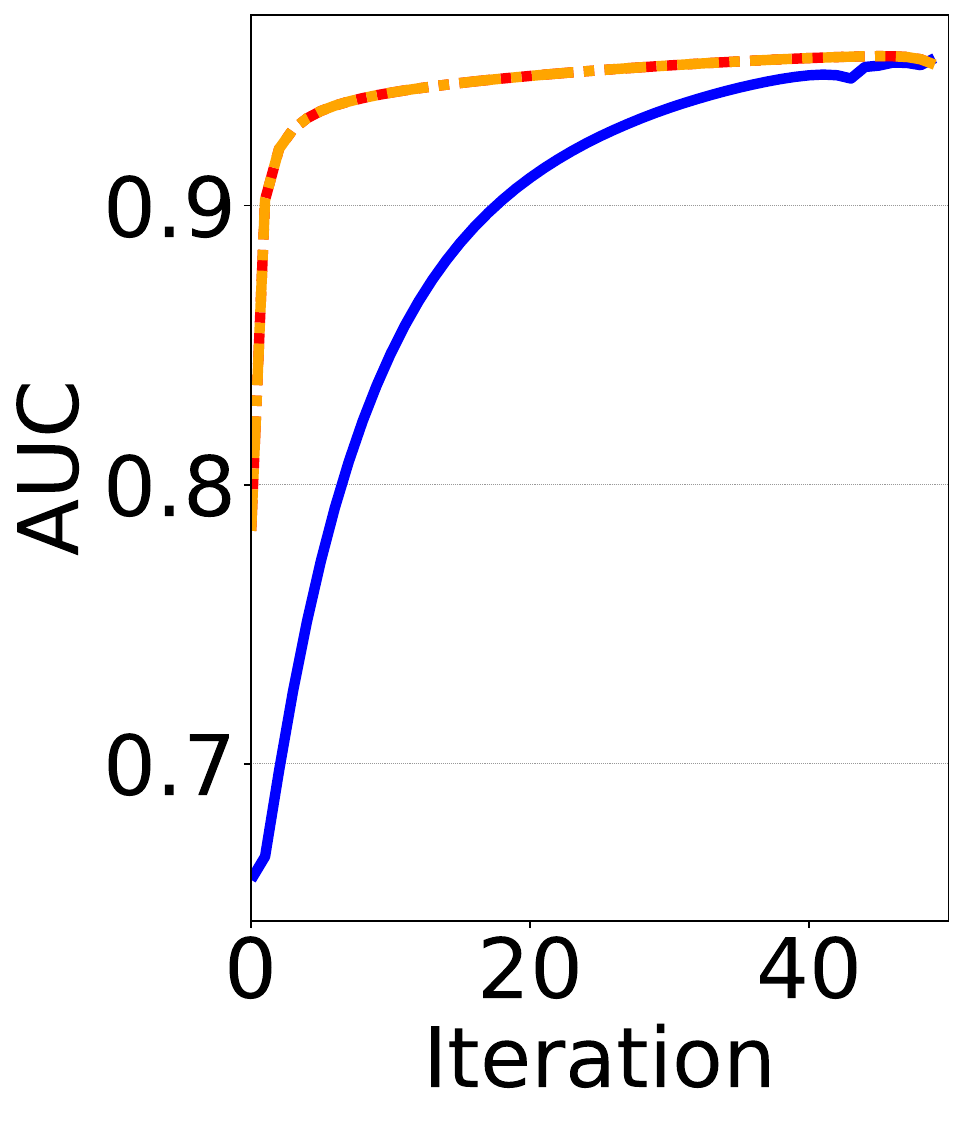}
		\subcaption{NUS-WIDE}
	\end{minipage} 
	\caption{Effect of PCA: AUC vs Iteration. Origin means vanilla VFL. PCA converts the input dimension to the target compressed dimension.}
	\label{fig-pca-performance-step}
\end{figure}

%% file: fig-pca-performance-time.tex
\begin{figure}[t]	
	\center		
	\captionsetup{justification=centering}	
	\begin{minipage}[b]{0.15\textwidth}
		\includegraphics[width=\textwidth]{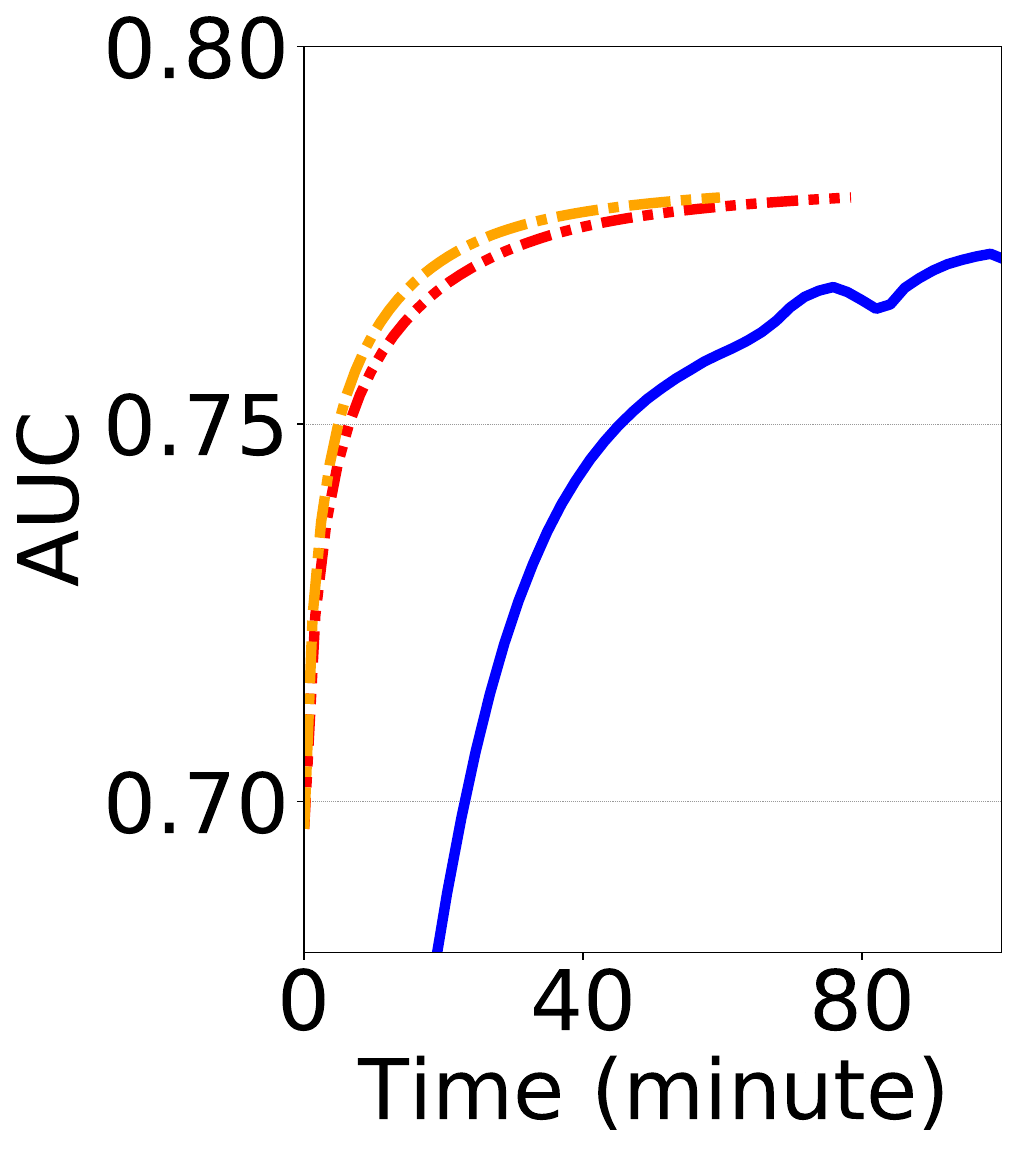}
		\subcaption{MIMIC}
	\end{minipage}
    ~
	\begin{minipage}[b]{0.15\textwidth}
		\hspace*{-35pt}\includegraphics[width=2.12\textwidth]{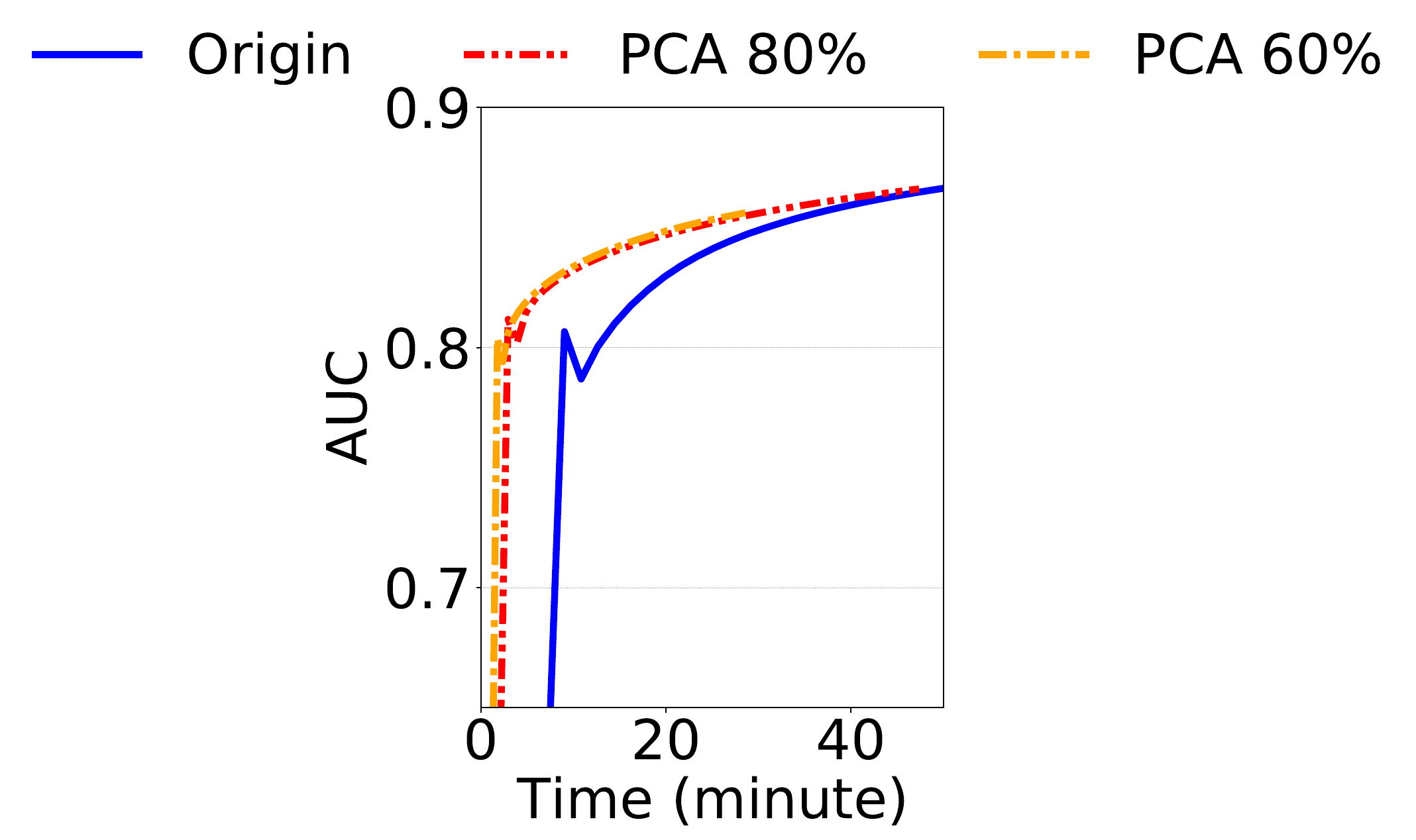}
		\subcaption{Epsilon}
	\end{minipage} 
	~
	\begin{minipage}[b]{0.145\textwidth}
		\includegraphics[width=\textwidth]{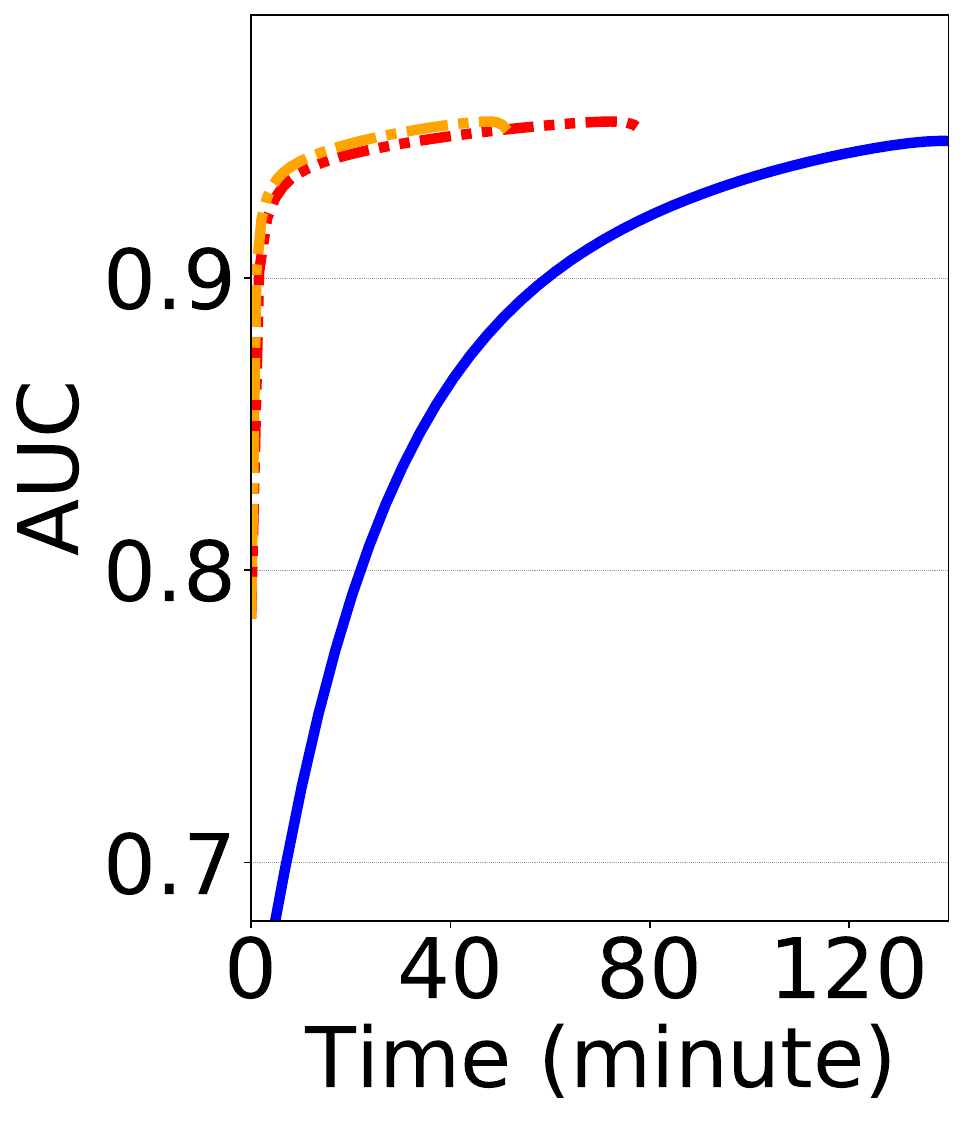}
		\subcaption{NUS-WIDE}
	\end{minipage} 
	\caption{Runtime performance of PCA: AUC vs Time.}
	\label{fig-pca-performance-time}
	
\end{figure}

%% file: tab-performance.tex
    

\begin{table}[]
    \normalsize
    \centering
    \begin{tabular}{cc|c|c|c|c}
    \hline
    \multicolumn{2}{c|}{Methods}                                                                             & Origin & Backup & PCA   & Ours  \\ \hline
    \multicolumn{1}{l|}{\multirow{3}{*}{\begin{tabular}[c]{@{}c@{}}Time\\ (minute)\end{tabular}}} & Comp. & 98.2    & 93.6   & 58.5  & 59.9  \\ \cline{2-6} 
    \multicolumn{1}{l|}{}                                                                         & Comm. & 141.0   & 48.3   & 137.6 & 46.4  \\ \cline{2-6} 
    \multicolumn{1}{l|}{}                                                                         & Sum   & 239.2   & 141.9  & 196.1 & 106.3 \\ \hline
    \end{tabular}
    \caption{Runtime performance under different modes. Origin means vanilla VFL without any optimization. Number of backup workers is set to 2 in Backup mode. PCA mode reduces the feature dimension to 60\%. Ours means the whole accelerating system composed of both Backup and PCA. All the four methods are carried out for 50 iterations on \texttt{MIMIC-III}.}
    \label{tab-mixture}
    \end{table}

%% file: related.tex
\section{Related Work}
\label{related}
\subsection{Related Framework}
In order to get rid of the computation and communication redundancy caused by homomorphic encryption, 
Hu et.al propose a novel vertical federated learning framework on the assumption of shared labels~\cite{Hu2019}. This assumption may be valid within a 
single cilo. But when the scenario is extended to cross-cilo cooperation, many labels shall not be shared due to privacy regulation and corporate competition, 
limiting it to be applied in large-scale industrial scenarios. 
Therefore, to the best of our knowledge, VFL under homomorphic encryption is still the safest framework for cross-silo cooperating training except from multi-party secure computing~\cite{Zheng2021}, the efficiency of which is far from usability in practical.
Some works have been proposed to make homomorphic encryption more suitable for federated learning.
In the horizontal scenario, Wang et.al designed a new coding scheme 
for the homomorphic encryption scheme~\cite{Zhang2020}, which greatly reduces the redundancy of the homomorphic encryption scheme. However, the 
assumption that all parties can decrypt data and calculate in local plaintext is difficult to hold in vertical federated 
learning. Therefore, it can not solve the huge bottleneck of encrypted computation in vertical federated learning. 


\subsection{Network heterogeneity and computing heterogeneity}


Heterogeneity is the main bottleneck of cooperative learning. Conventional distributed learning adopts many efficient schemes to 
eliminate the overhead caused by heterogeneity.
One traditional way is to reduce the impact of heterogeneity by increasing the degree of parallelism, such as pipeline~\cite{Bai2020}, or selecting those similar clients or data to prevent heterogeneity ahead~\cite{4}.
Asynchronous training~\cite{Lian2017} is another common paradigm. It solves the heterogeneity problem at 
the cost of introducing the staleness, which damages the convergence performance of training. However, in the typical VFL framework, jobs are limited to be trained in a synchronized way. So common asynchronous optimization comes out of effect.
A new paradigm called backup worker was introduced in~\cite{Chen2016}. Assuming there is $n+b$ worker, backup scheme makes aggregator get the first $n$ gradients and drop the other $b$ gradients as backup.
While in vertical federated learning, the parameters transmitted are not only the updated gradients, but also the encrypted intermediate results, which drag the communication the most. 

%% file: conclusion.tex
\section{Conclusion and Future Work}
\label{conclusion}
In this paper, we dive into the bottleneck of vertical federated learning under homomorphic encryption, and propose a practical and effective optimization system to accelerate current vertical federated learning through backup worker and principal component analysis. 
The backup scheme improves the 
efficiency and robustness of the communication network in the heterogeneous scenarios. Principal component analysis alleviates the computation overhead caused by homomorphic encryption.
Our system dramatically accelerates current vertical federated learning and promotes its practical deployment in safe cross-industry cooperation. 
We will further optimize the performance and efficiency of vertical federated learning in the future, mainly in the following four directions.

\subsection{AUTO-ML}
The underlying algorithm 
of FATE is completely redeveloped to ensure security. Hence, model hyper-parameters that perform well in traditional distributed learning can not be applied directly. During our experiments, it took us a lot of effort to adjust these hyper-parameters and keep the training performance of VFL almost consistent with local training.
Auto-ML~\cite{xu2020federated} can be used to automatically adjust parameters, so that the potential of VFL can be maximized.

\subsection{Encryption Optimization}
While reduction of computation overhead is significantly apparent, it is at the cost of possible accuracy loss, which is far from perfect. 
This tradeoff between computational efficiency and model accuracy is related to the tradeoff between privacy preserving and model accuracy for horizontal federated learning, which has been theoretically investigated by \cite{zhang2022no}.  
Considering the computation overhead in VFL is mainly caused by the multiplication of homomorphic encryption, we will further optimize the compression scheme through finding a multiplication-friendly encryption strategy to eliminate computation overhead in the current vertical federated learning without losing security.

\subsection{More datasets and metrics}

The existing vertical federated learning does not have a typical dataset or standard for people to refer. Lack of datasets and metrics forces researchers in VFL to segment and generate data sets according to their 
own understanding and compare them with themselves. However, this method of segmentation is probably unscientific, unprofessional, and impractical. A professional and practical vertical dataset provided by specialists in related fields will be of great help.


\subsection{Incentive mechanism}
The cooperators in VFL training need to have fruitful data features for effective collaborative training.
They tend to be competitive enterprises in certain fields, such as industry, finance, medicine, etc.
Attracting these enterprises to build collaboration is non-trivial.
The incentive mechanism design for VFL is a promising research topic to pave the way for practical VFL.
For example, rewarding the model owners based on their marginal contributions can ensure the stability of a federation~\cite{lim2020hierarchical}.
And how to measure the reliability and contributions of the cooperators is a sophisticated optimization problem as well~\cite{kang2019incentive}.
Those important topics researched in conventional FL are equally meaningful in the VFL scenario.
We will design a reasonable and efficient incentive mechanism for modern VFL in the future.

%% file: photos.tex
\begin{IEEEbiography}[{\includegraphics[width=1in,height=1.25in,clip,keepaspectratio]{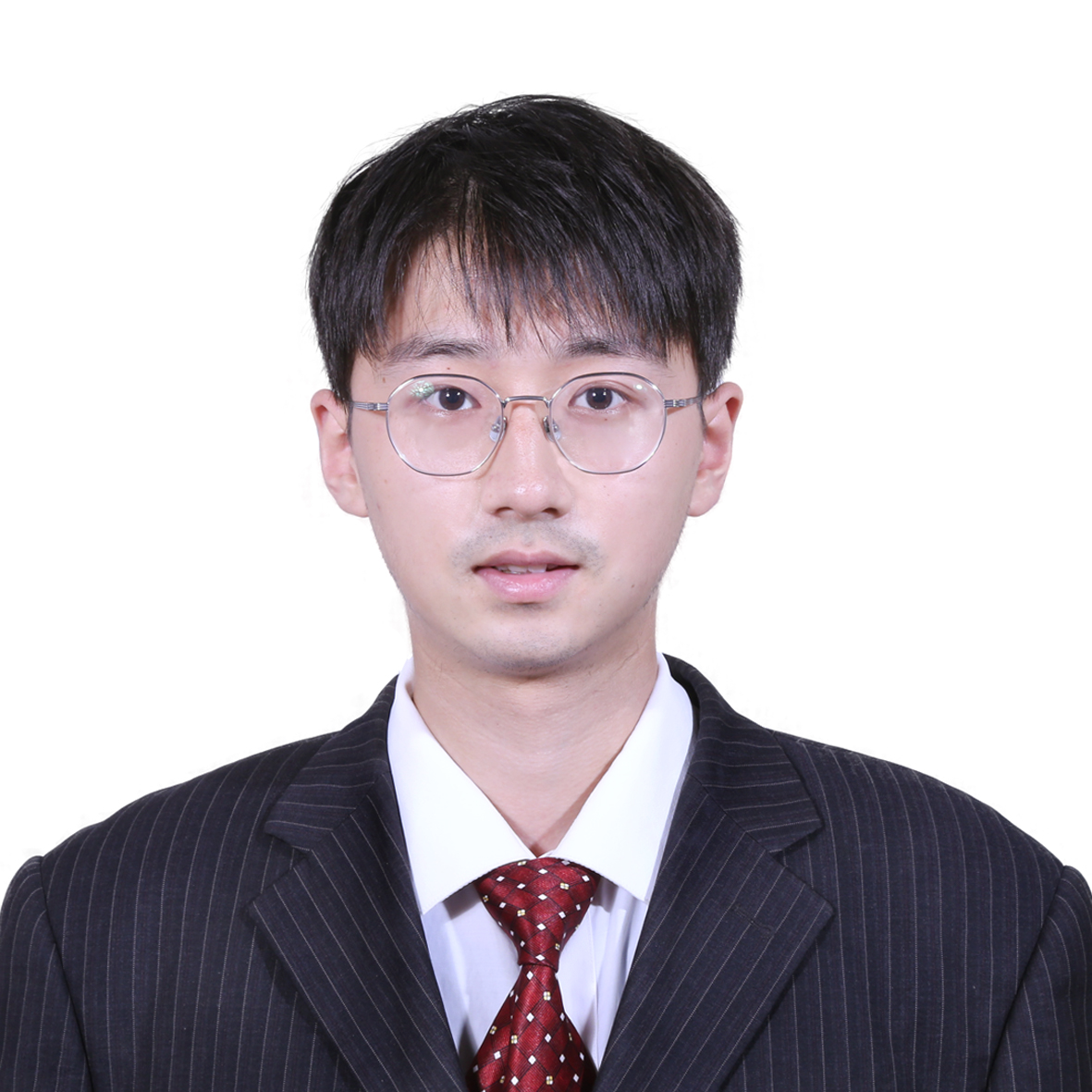}}]{Dongqi Cai}
    is currently a Ph.D. candidate at the State
Key Laboratory of Networking and Switching
Technology, Beijing University of Posts and
Telecommunications. His research
interests include mobile computing and federated learning.
\end{IEEEbiography}

\begin{IEEEbiography}[{\includegraphics[width=1in,height=1.25in,clip,keepaspectratio]{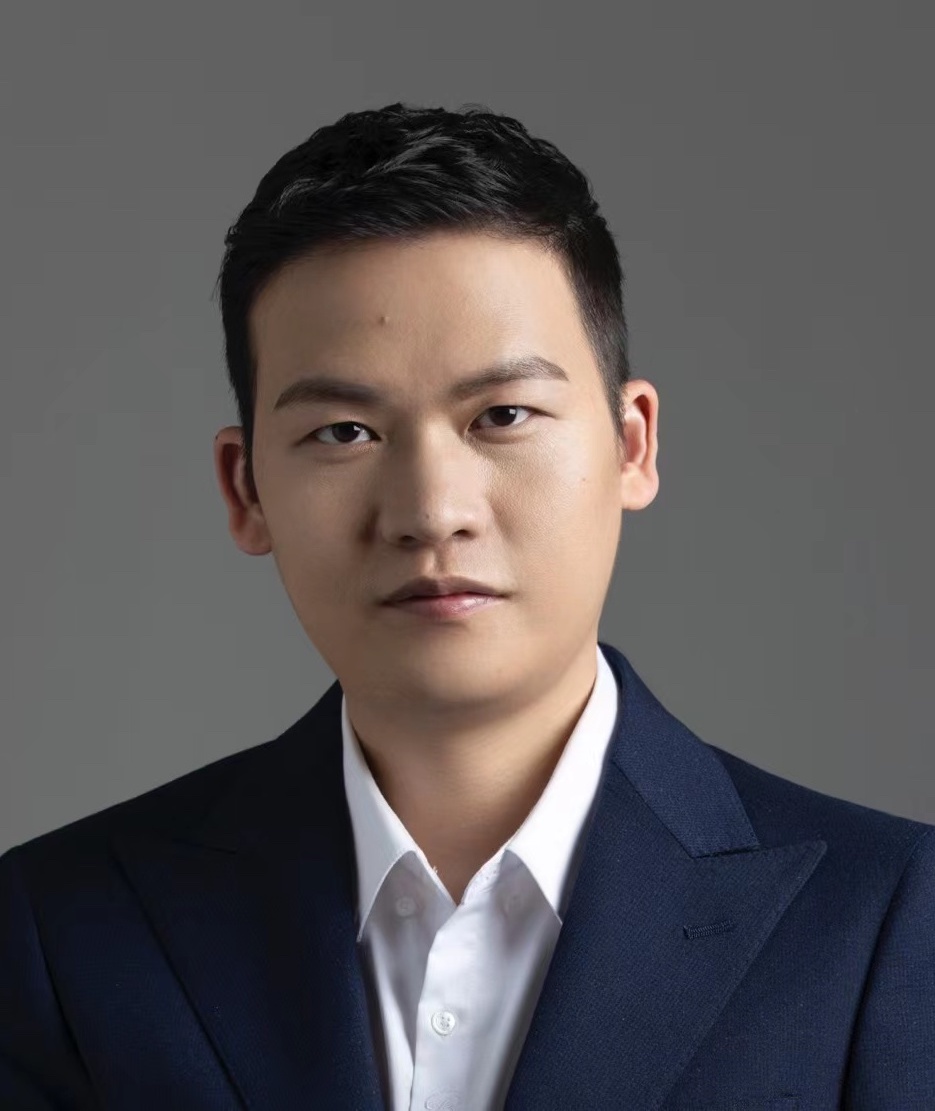}}]{Tao Fan}
    is a  Principal Researcher and Tech Lead in the AI Department of WeBank, ShenZhen, China. He is now responsible for FATE, an famous industrial level federated learning open source project. 
Before joining WeBank, he worked at Baidu, Tencent. He has more than 8 years of experience in large-scale machine learning system and  big data application fields. He received his Master degree from University of Science and Technology of China in 2013 and his Bachelor's degree from Shandong University of Science and Technology in 2010. 
\end{IEEEbiography}

\begin{IEEEbiography}[{\includegraphics[width=1in,height=1.25in,clip,keepaspectratio]{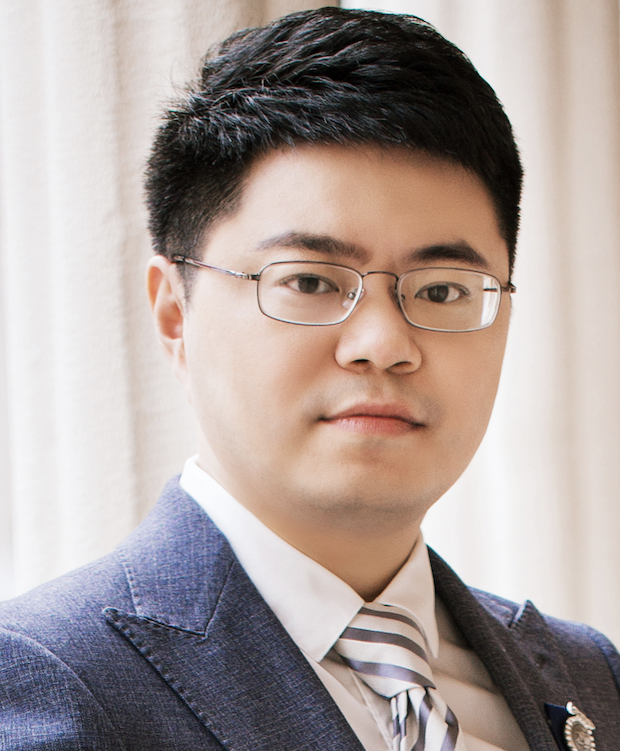}}]{Yan Kang}
    is currently a research team lead with the AI Department of WeBank,, Shenzhen, China. His works focus on the research and implementation of privacy-preserving machine learning and federated learning. His research was authored or coauthored in well-known conferences and journals including IEEE Intelligence Systems, IJCAI, and ACM TIST, and coauthored the Federated Learning book.
\end{IEEEbiography}

\begin{IEEEbiography}[{\includegraphics[width=1in,height=1.25in,clip,keepaspectratio]{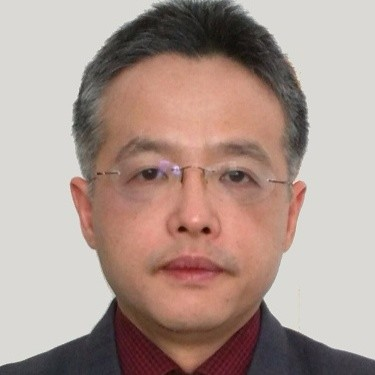}}]{Lixin Fan}
    is the Principal Scientist of Artificial Intelligence at WeBank and the Chairman of the Federal Learning Industry Ecological Development Alliance. His research fields include machine learning and deep learning, computer vision and pattern recognition, image and video processing, 3D big data processing, data visualization and rendering, augmented and
virtual reality, mobile computing and ubiquitous computing, and intelligent man-machine interface. He is the author of more than 60 international journals and conference articles and has received more than 7,000 citations. He has worked at Nokia Research Center and Xerox Research Center Europe. His research includes the well-known Bag of Keypoints image classification method. He has participated in NIPS/NeurIPS, ICML, CVPR, ICCV, ECCV, IJCAI and other top artificial intelligence conferences for a long time, served as area chair of AAAI, and organized workshops in various technical fields. He is also the inventor of nearly one hundred patents filed in the United States, Europe and China, and the chairman of the IEEE P2894 Explainable Artificial Intelligence (XAI) Standard Working Group.
\end{IEEEbiography}

\begin{IEEEbiography}[{\includegraphics[width=1in,height=1.25in,clip,keepaspectratio]{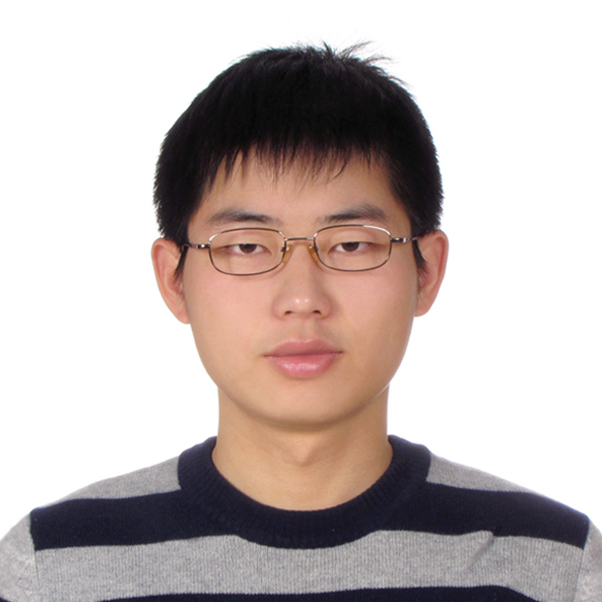}}]{Mengwei Xu}
    is an Assistant Professor with the Beijing
University of Posts and Telecommunications, Beijing, China.
His research interests include mobile and edge computing.
He received the B.S. and Ph.D. degrees from Peking University, Beijing, China.
\end{IEEEbiography}

\begin{IEEEbiography}[{\includegraphics[width=1in,height=1.25in,clip,keepaspectratio]{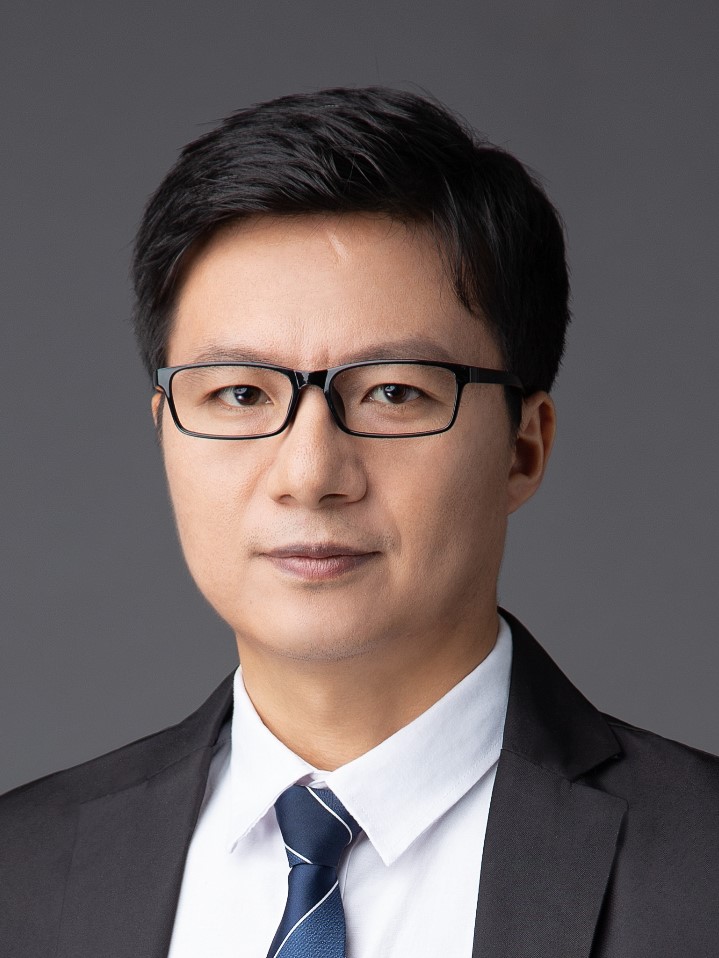}}]{Shangguang Wang}
    is a Professor at the School of Computer Science and Engineering, Beijing University of Posts and Telecommunications, China. He received his Ph.D. degree at Beijing University of Posts and Telecommunications in 2011. He has published more than 150 papers. His research interests include service computing, mobile edge computing, and satellite computing. He is currently serving as Chair of IEEE Technical Committee on Services Computing (2022-2023), and Vice-Chair of IEEE Technical Committee on Cloud Computing (2020-). He also served as General Chairs or Program Chairs of 10+ IEEE conferences. He is a Fellow of the IET, and Senior Member of the IEEE. For further information on Dr. Wang, please visit: http://www.sguangwang.com 
\end{IEEEbiography}

\begin{IEEEbiography}[{\includegraphics[width=1in,height=1.25in,clip,keepaspectratio]{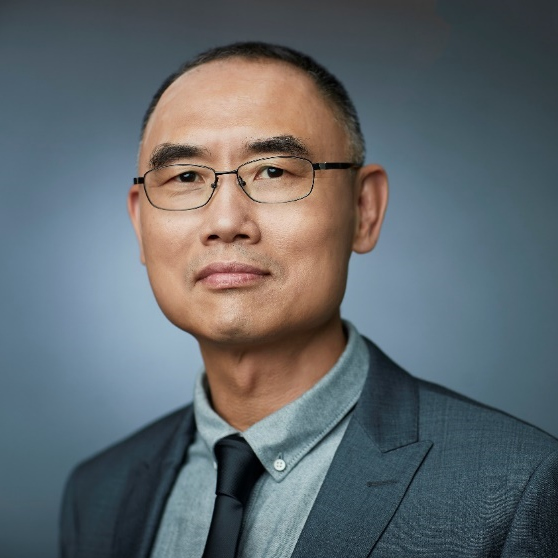}}]{Qiang Yang}
    is Chief Artificial Intelligence Officer of WeBank and Chair Professor of CSE Department
of Hong Kong Univ. of Sci. and Tech. He is the Conference Chair of AAAI-21, President of Hong Kong Society of Artificial Intelligence and Robotics (HKSAIR) , the President of Investment Technology League (ITL) and a former President of IJCAI (2017-2019). He is a fellow of AAAI, ACM, IEEE and AAAS. His research interests include transfer learning and federated learning. He is the founding EiC of two journals: IEEE Transactions on Big Data and ACM Transactions on Intelligent Systems and Technology.
\end{IEEEbiography}